%% file: Manuscript.tex
\documentclass[smallcondensed]{svjour3}     % onecolumn (ditto)
\smartqed  % flush right qed marks, e.g. at end of proof
\usepackage{graphicx}
\journalname{Empirical Software Engineering}

\input{packages}

\begin{document}
\setcounter{tocdepth}{2}
\setcounter{secnumdepth}{3}

\input{commands}

\title{Using Small Language Models to Reverse-Engineer Machine Learning Pipelines Structures
  \thanks{This work is funded by the FATES-MLOps international cooperation project (ANR (TSIA 2024) ANR-24-IAS2-0002-01, NSERC – ALLRP-2024-599951) and by the French National Research Agency (ANR) under grant agreement no. ANR-24-CE25-1286-01.}
}
\subtitle{Potential and Limits}

\titlerunning{Using SLMs to Reverse-Engineer ML Pipelines Structures}        % if too long for running head

\author{
  Nicolas Lacroix\orcidlink{0009-0004-4390-6694}
  \and
  Frédéric Precioso\orcidlink{0000-0001-8712-1443}
  \and
  Mireille Blay-Fornarino\orcidlink{0000-0001-9852-7764}
  \and
  Sébastien Mosser\orcidlink{0000-0001-9769-216X}
}

\authorrunning{Nicolas Lacroix\and Frédéric Precioso \and Mireille Blay-Fornarino \and Sébastien Mosser} % if too long for running head

\institute{
  Nicolas Lacroix*\and Frédéric Precioso \at
  Université Côte d'Azur, Inria, CNRS, I3S, Sophia Antipolis, France\\
  \email{nicolas.lacroix@univ-cotedazur.fr}\\
  \email{frederic.precioso@univ-cotedazur.fr}
  \and
  Mireille Blay-Fornarino \at
  Université Côte d'Azur, CNRS, I3S, Sophia Antipolis, France\\
  \email{mireille.blay@univ-cotedazur.fr}
  \and
  Sébastien Mosser \at
  McSCert, McMaster University, Hamilton, ON, Canada\\
  \email{mossers@mcmaster.ca} \\
  *Corresponding author.
}

\date{Received: date / Accepted: date}

\maketitle

\input{0_Abstract}
\input{1_introduction}
\input{2_related_work}
\input{3_reproduction}
\input{4_study_design}
\input{5_study_results}
\input{6_deviations_registered_report}
\input{7_threats_to_validity}
\input{8_conclusion}
\input{9_declarations}
\input{10_appendix}

\bibliographystyle{IEEEtran}
\bibliography{references}

\end{document}

%% file: packages.tex
\usepackage{cite}
\usepackage{amsmath,amssymb,amsfonts}
\usepackage{algorithmic}
\usepackage{graphicx}
\usepackage{orcidlink}
\usepackage{xspace}
\usepackage{listings}
\usepackage{xargs}
\usepackage[table,dvipsnames]{xcolor}
\usepackage[most]{tcolorbox}
\usepackage{booktabs}
\usepackage{url}
\usepackage{rotating}
\usepackage{makecell}
\usepackage{adjustbox}
\usepackage{placeins}
\usepackage{hyperref}

%% file: commands.tex
\newcommand{\eg}{\emph{e.g.},\xspace}
\newcommand{\ie}{\emph{i.e.},\xspace}
\newcommand{\wrt}{\emph{w.r.t.}\xspace}
\newcommand{\cf}{\emph{cf.}\xspace}
\newcommand{\etal}{et al.\xspace}
\newcommand{\aka}{\emph{a.k.a.},\xspace}

\newcommand{\taxounified}{\textit{T\textsubscript{unified}}\xspace}
\newcommand{\taxodspipelines}{\textit{T\textsubscript{dspipelines}}\xspace}
\newcommand{\taxodaswow}{\textit{T\textsubscript{daswow}}\xspace}

\newcommand{\datasetdspipelines}{\textit{D\textsubscript{dspipelines}}\xspace}
\newcommand{\datasetdaswow}{\textit{D\textsubscript{daswow}}\xspace}
\newcommand{\datasetrule}{\textit{D\textsubscript{rule}}\xspace}

\newcommand{\expert}[1]{\textit{E\textsubscript{#1}}}

\newcommand{\slmbest}{\textit{SLM\textsubscript{best}}\xspace}
\newcommand{\cslacroix}{\textit{CS\textsubscript{lacroix}}\xspace}

\tcbset{
  soberresult/.style={
    boxrule=0.5pt,
    arc=2pt,
    left=6pt, right=6pt, top=6pt, bottom=6pt,
    fonttitle=\bfseries,
    % coltitle=black,
    title=Finding,
  }
}

\definecolor{quotebar}{HTML}{D0D7DE}
\definecolor{quotebg}{HTML}{F6F8FA}
\newtcolorbox{markdownquote}{
    enhanced,
    breakable,
    colback=quotebg,
    colframe=quotebar,
    boxrule=0pt,
    leftrule=2.5pt,
    arc=0pt,
    left=8pt,
    right=6pt,
    top=4pt,
    bottom=4pt,
    before skip=8pt,
    after skip=8pt,
}

\lstset{
  columns=fullflexible,
  frame=single,
  breaklines=true,
}

%% file: 0_Abstract.tex
\begin{abstract}
  \textit{Context}: Once defined a taxonomy of stages structuring Machine Learning (ML) pipelines (e.g. Data Preprocessing, Modeling...), extracting these stages from source code is key for better understanding ML practices. However, the diversity caused by the constant evolution of ML (\eg~algorithms, datasets) makes this task challenging. Existing approaches either rely on non-scalable manual labeling or on classifiers that do not properly support domain's diversity. These limitations call for more reliable solutions. \textit{Objective}: We evaluate whether Small Language Models (SLMs) can leverage their code understanding and classification abilities to address these limitations, and enhance our understanding of practices in ML. \textit{Method}: We conduct a confirmatory study based on two relevant reference works representing current limitations in the state-of-the-art. We first compare several SLMs using Cochran's Q test, then evaluate the best-performing model against reference studies via two McNemar's tests. An additional Cochran's Q test examines how taxonomy definition variations affect the SLM performance. Finally, goodness-of-fit tests compare ML practice insights from SLM classification with those from prior studies. \textit{Results}: First, we found that the taxonomy wording significantly impacts classification performance. Second, the best performing SLM yielded good results, yet, without outperforming other classifiers. Third, the three classification methods led to significantly different insights, with varying effect sizes, when exploring practices of data scientists. \textit{Conclusions}: Limitations of existing classification methods bias our understanding of ML practices. While current SLMs show promising results without prior fine-tuning, they still exhibit common limitations, in addition to inference high costs challenging their applicability in large-scale studies.
  \keywords{Reverse Engineering \and Machine Learning Pipelines \and Small Language Models \and Jupyter Notebooks}
\end{abstract}

%% file: 1_introduction.tex
\section{Introduction}
\label{s:intro}

Over the last two decades, Machine Learning (ML) got a pervasive impact, solving increasingly complex decision tasks, owing to new families of methods every two or three years. ML democratization lead to a huge and ever increasing amount of contributions (i.e. ML workflows/pipelines), of various qualities, exploring to address new more complex tasks.  
Understanding how data scientists structure their ML pipelines to answer a problem, is key to better supporting the quality and evolution of these pipelines. Studying them at an abstract level enables the analysis of how workflows are organized, iteratively built, and reused for a new problem. This higher-level perspective is therefore crucial to characterize current practices empirically.

Recently, studies have examined pipeline quality~\cite{wang2025machine,islam2019comprehensive}, identified recurring patterns~\cite{bogner2021characterizing}, and supported practitioners through pipeline reuse~\cite{redyuk2024assisted}. All these efforts rely on accurately extracting the underlying structure of ML pipelines from source code. However, this task is often achieved manually because of the high diversity of the ML domain. Indeed, the exploratory nature~\cite{martinez2019crisp} of ML associated with the rapid evolution of new algorithm families and ML libraries has led to increasingly heterogeneous ML pipelines~\cite{biswas2022art,ramasamy2023workflow}.

From reliance on static mappings between specific libraries and pipeline stages \cite{jiang2022elevating,biswas2022art,venkatesh2023enhancing}, efforts have recently shifted toward automated approaches using rule-based or supervised classifiers~\cite{wang2021graph,zhang2022coral,perez2024flexible}. Despite their promising results, existing methods remain constrained in performance (\eg classification accuracy) and scope (\eg specific libraries or pipeline patterns). Furthermore, they often use different taxonomies of the different ML stages (\eg Data Preprocessing, Modeling, Evaluation...) with variations in their formulations. Consequently, the same limitations affect the empirical analysis of data science pipelines in practice.

Simultaneously, language models have shown strong capabilities in code understanding and classification tasks~\cite{hou2024large}. In addition, Small Language Models (SLMs) have drawn interest from the scientific community, looking for more sustainable and reproducible practices, thanks to their lower resource demands. Yet, their potential for extracting ML pipeline structures has still to be evaluated.

\textbf{Following our registered report}~\cite{lacroix2026using}, we present the results of our confirmatory study which investigates the use of SLMs for reverse-engineering ML pipeline structures, its limitations, and its impact on our understanding of ML practices. This paper addresses the following research questions:

\begin{description}
    \item{\textbf{RQ1}} In ML pipeline's structure extraction, are SLMs significantly more accurate in identifying stages than existing model-based or rule-based classifiers?
    \item{\textbf{RQ2}} Does the taxonomy ``wording'' significantly influence the classification results of SLMs?
    \item{\textbf{RQ3}} Does the use of an SLM in the reverse-engineering process provide significantly different exploratory insights than those produced by reference studies compared to the ground truth?
\end{description}

To answer \textbf{RQ1}, we selected two reference studies from the literature, and replicated their classification methods using provided reproduction packages. Building on their datasets, we benchmarked and selected the best performing SLM among a set of open-weight SLMs trained for coding tasks. Then, we created a mapping between the different reference studies' taxonomies to obtain a unified taxonomy. This taxonomy was used to compare classification results from this SLM with the reference studies, both from a statistical and practical perspective.

To answer \textbf{RQ2}, we mutated the unified taxonomy's headwords using the ``synonym substitution''~\cite{chen2026nlperturbator} method. Each mutation impact was evaluated using the same benchmarking methodology that RQ1, with the SLM being prompted with the mutated taxonomy.

To answer \textbf{RQ3}, we analyzed key insights (\eg~stages distribution, transition frequencies, sequential patterns) emerging from the classification results obtained using the reference studies and the best performing SLM. These insights were compared according to the classification method used with manually labeled ML pipelines to assess the ability of these classification methods to provide a consistent understanding of the practices in ML.

\textbf{Contribution} First, we replicate two scientific studies and show empirically that our understanding of practices in ML is influenced by limitations in the classification methods currently used in the literature. This demonstrates that there remains a significant amount of work to be done to develop a comprehensive understanding of these practices. We also show that available SLMs are not an immediate solution to this problem. Despite offering promising results, we demonstrate that they are highly sensitive to the formulation of the taxonomy used. This is particularly important as the literature presents a multitude of taxonomies. Finally, our contribution also lies in the adoption of a more sustainable scientific approach (better reproducibility and contained energy consumption with smaller language models), as motivated by ``The Copenhagen Manifesto''~\cite{russo2024generative}. Our study demonstrates that SLMs still present a high inference cost, a factor that must be considered before undertaking costly fine-tuning or turning to larger models.

\begin{figure}[t]
\caption{Overview of the main tasks of our study}
\label{fig:study_components_overview}
\includegraphics[width=11cm]{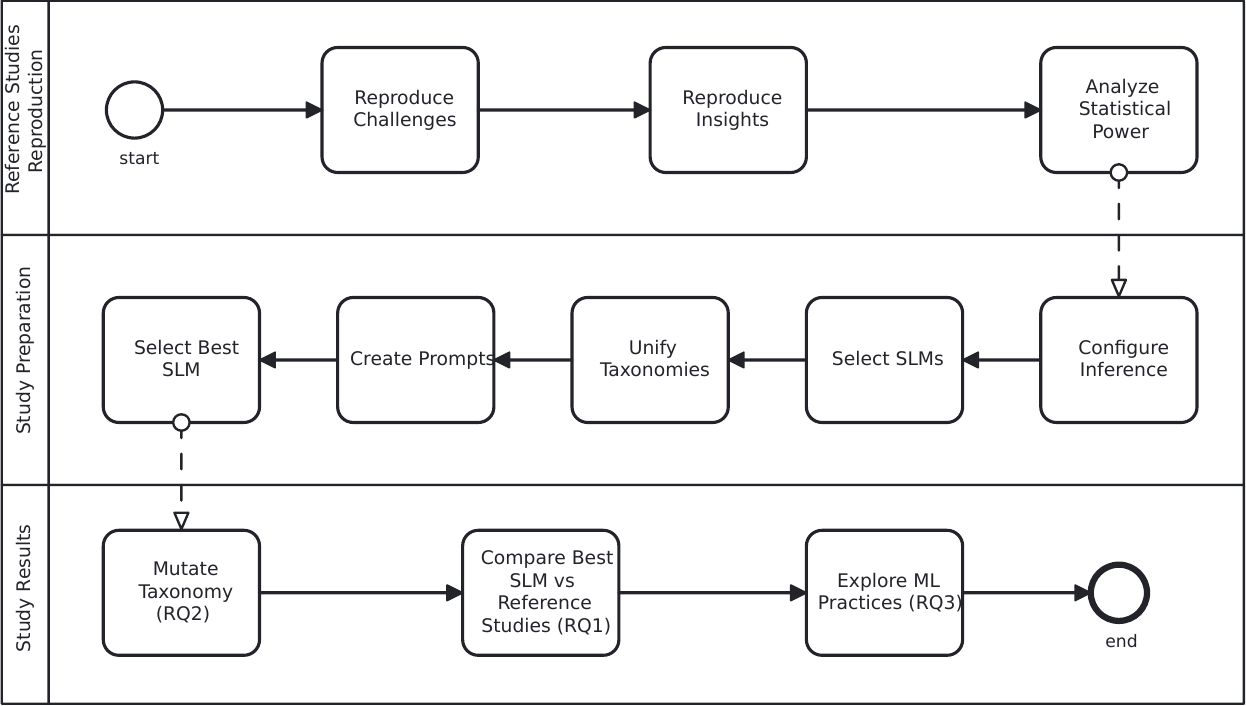}
\centering
\end{figure}

\textbf{Paper structure} The overview of our study is illustrated in Figure~\ref{fig:study_components_overview}. We provide background and discuss related work of the study in Section~\ref{s:background_related_work}. We present key insights from the reference studies and discuss the challenges encountered in replicating them in Section~\ref{s:reference_studies_reproduction}. We introduce the preparation of the study, including the best performing SLM selection and taxonomy unification, in Section~\ref{s:study_preparation}. We introduce the study results and answer research questions in Section~\ref{s:study_results}. We clarify the few deviations from the registered report in Section~\ref{s:deviations_registered_report}. We discuss the threats to validity in Section~\ref{s:threats_validity}. We conclude and discuss future works in Section~\ref{s:conclusion_future_works}.

%% file: 2_related_work.tex
\section{Background and Related Work}
\label{s:background_related_work}

\subsection{Reverse-Engineering and Language Models}
\label{ss:llm_reverse_engineering_tasks}

While code summarizing and documentation generation remain very common applications~\cite{hou2024large}, researchers have recently become interested in the ability of language models to extract structured information from code bases. More specifically, Boronat et al.~\cite{boronat2025mdre} have proposed a tool, \textit{MDRE-LLM}, that integrates LLMs with traditional MDRE techniques for extracting high-level abstractions from source code. Additionally, Siala et al.~\cite{siala2025comparison} studied the use of LLMs for reverse-engineering Python and Java code to the OCL formal language. Similarly, Campanello et al.~\cite{campanello2025use} have shown promising results in the use of LLM (GPT-4) for reverse-engineering class (as PlantUML) diagrams. However, the literature mainly focuses on using \textit{large} language models, which presents a high inference cost and limitations to the reproducibility of scientific experiments (requires access to computing infrastructures or the use of gated models).

In the same time, recent works have demonstrated the potential of using smaller language models, as illustrated by REx86\cite{lea2025rex86} (a fine-tuned version of Qwen2.5-Coder-7B) for x86 Assembly. Similarly, Ouf et al.\cite{ouf2025reverse} have shown that smaller language models ($8$B parameters) can match larger language models performance ($70$B parameters) in reverse-engineering code with specific prompting techniques. More importantly, the fact that these SLMs have not undergone fine-tuning suggests that they can achieve strong performance without incurring the additional cost of such fine-tuning. Accordingly, and following recommendations made by Wagner et al.~\cite{wagner2025towards} (open models, transparent prompts and models configuration), we address the need to reverse-engineer ML pipelines by investigating the use of SLMs, which has not been previously examined.

\subsection{From Structured Frameworks to Exploratory Practices}
\label{ss:llm_code_analysis}

The CRISP-DM framework~\cite{DMwirth2000crisp} remains the \textit{de facto} historical standard for modeling the Data Mining process, introducing structured notions of sequential tasks (\eg~``Data Preparation'', ``Modeling'') and outputs. However, the shift over the last decade from purely descriptive Data Mining to the predictive and prescriptive domain of Data Science has increased the reliance on Machine Learning (ML) models and domain expertise, rendering these frameworks often inadequate for current exploratory practices~\cite{martinez2019crisp}. Indeed, the adoption of literate programming tools, most notably Jupyter Notebooks, has enabled practitioners to quickly experiment and iterate throughout the entire pipeline lifecycle, from data preparation to model evaluation. This freedom has led to a massive increase in the volume of ML pipelines developed. According to the Octoverse $2025$ report\footnote{\resizebox{0.95\textwidth}{!}{\url{https://github.blog/news-insights/octoverse/octoverse-a-new-developer-joins-github-every-second-as-ai-leads-typescript-to-1/}}}, the amount of GitHub repositories using notebooks has increased by $+75\%$ compared to $2024$, resulting in a total of $2.4$ million repositories in $2025$. More critically, the variability and heterogeneity of these pipelines~\cite{lau2020design,brault2023taming} directly complicate the automatic extraction and identification of pipeline structures, which are essential prerequisites for large-scale empirical analysis.

\subsection{Reverse-Engineering Applied to ML Pipelines Extraction}
\label{ss:auto_pipeline_extract}

Software engineering researchers have a long-standing interest in analyzing scientific workflows~\cite{stevens2001classification,wassink2009analysing} and managing computational processes~\cite{garijo2013detecting,liu2015survey,patterson2017dataflow}, with a growing focus on the extraction of ML pipeline structures from source code for practices understanding and evaluation.

Static mapping or rule-based approaches identify stages by matching them to predefined functions or libraries. For instance, initial work by Venkatesh et al.~\cite{venkatesh2023enhancing} identified stages based on a static mapping of $13$ libraries, integrating these as documentation. Similarly, Jiang et al.~\cite{jiang2022elevating} proposed around $700$ hard-coded rules derived from a few selected ML libraries and have shown that one cell can contain code instructions related to multiple stages. Biswas et al.~\cite{biswas2022art} used a static mapping composed of $403$ specific functions to explore the practices of data scientists, giving first insights about stages frequency and ordering in Kaggle notebooks.

Works leveraging supervised classifiers have tried to mitigate these static limitations. For example, Zhang et al.'s CORAL solution~\cite{zhang2022coral} uses an ML classifier trained on a custom taxonomy but is restricted to seven Data Science libraries (pandas, statsmodels, etc.). More recently, Venkatesh et al. has extended their initial work and proposed a supervised classifier~\cite{venkatesh2024static} trained on their previous mapping of $13$ libraries. Similarly, Perez et al.~\cite{perez2024flexible} tried to improve this approach by combining a decision tree model with a rule-based model. In their comprehensive empirical study, Ramasamy et al.~\cite{ramasamy2023workflow} also developed a classifier and explored the practices of data scientists. They confirmed the imbalanced stage frequencies and transition probabilities, but showed mixed results in their classifier evaluation (e.g., poor \textit{F}1 score of $0.29$ for prediction), highlighting the need for more flexible reverse-engineering solutions.

Overall, these approaches have demonstrated promising results within specific contexts. However, their reliance on predefined rules or narrowly trained datasets limits their scalability and generalization to the highly variable ML ecosystem. Instead, we propose to benefit from the ability of language models to learn statistical patterns, which frequently arise in the code~\cite{hindle2016naturalness}. Additionally, these models are trained on extensive datasets containing billions of code files (\eg \textit{The Stack v2}\cite{lozhkov2024starcoder}). These include the source code for millions of ML pipelines mined from popular hosting platforms (\eg Kaggle, Github), reflecting the diversity of the field. As a result, we expect language models to better support the variability of the ML domain.

%% file: 3_reproduction.tex
\section{Reference Studies and Reproduction}
\label{s:reference_studies_reproduction}

In this section, we detail the reference studies selection and replication process. We also discuss its limitations and present some interesting insights about both reference studies' models and datasets.

\subsection{Reference Studies Selection}
\label{ss:reference_studies_selection}

Reference studies were selected according to the following criteria:

\begin{enumerate}
    \item \textit{C\textsubscript{1}}: the study reverse-engineers ML pipeline structures (ML stages) from code
    \item \textit{C\textsubscript{2}}: the study presents empirical results on the exploration of data scientist practices, necessary for comparison in \textbf{RQ3}
    \item \textit{C\textsubscript{3}}: the study provides an executable, open-source reproduction package
\end{enumerate}

After selecting works matching \textit{C\textsubscript{1}} from the literature, we filtered out works either violating \textit{C\textsubscript{2}} (\cite{wang2022documentation,wang2021graph,patterson2017dataflow,zhang2022coral,garijo2013detecting}) or \textit{C\textsubscript{3}} (\cite{perez2024flexible,venkatesh2024static,venkatesh2023enhancing}). This left us with three studies \cite{ramasamy2023workflow,biswas2022art,jiang2022elevating} that we compared based on their \textit{depth} (single instruction vs. entire notebook cell classification), \textit{breadth} (library coverage), and \textit{publication date}. The first criterion relates to the level of granularity proposed by the different studies, which can impact further analysis (\eg we cannot retrieve the exact stages ordering if one cell involves multiple stages). The second criterion involves the ability to extend to new libraries and algorithms, which is an important limitation identified in the literature\cite{venkatesh2024static}. The third criterion ensures the selected works remain relevant at the time of our study. Given these criterion, the study \textit{DS-Pipelines} stood out as the only one to classify code at the instruction level, while \textit{DASWOW} was the most recent study (compared to \cite{jiang2022elevating}) to avoid explicitly relying on a subset of libraries. As a result, these two noteworthy studies form the reference baselines for our confirmatory study:

\begin{itemize}
  \item \textit{DS-Pipelines\footnote{\url{https://github.com/sumonbis/DS-Pipeline}}}~\cite{biswas2022art}: works at the instruction level and relies on a static mapping (rule-based classifier), created manually.
  \item \textit{DASWOW\footnote{\url{https://zenodo.org/records/7109939}}}~\cite{ramasamy2023workflow}: works at the cell level and relies on a supervised classifier (Random Forest Classifier)
\end{itemize}

We confirmed our ability to conduct our study by reproducing the results of both studies from their reproduction packages. These reproductions are provided as part of our reproduction package. We detail below some of the challenges we faced during this reproducibility process and insights gained from the analysis of their datasets.

\subsection{DS-Pipelines}
\label{ss:reproduction_dspipelines}

The \textit{DS-Pipelines} study is divided into three parts. First, they analyzed theoretical pipelines presented in the literature. Second, they automatically analyzed Jupyter notebooks, which are a popular support to construct such pipelines~\cite{granger2021jupyter}. Third, they manually analyzed open-source data science projects from GitHub. In our study, we address the second part, allowing for a more systematic classification thanks to their linear structure. Accordingly, the \textit{DS-Pipelines} raw dataset contains $105$ notebooks and a static mapping file \textit{stages.csv}. This file maps $404$ functions or classes names to a stage (encoded as a number). Each stage is defined as part of a taxonomy \taxodspipelines of $11$ stages (\cf Table~\ref{tab:dspipelines_taxonomy} in the Appendix), with only $6$ of them being used when analysing notebooks. The result dataset,~\datasetdspipelines, is obtained by applying this mapping to each notebook instruction using their open-source parser.

\subsubsection{Challenges}
\label{ss:reproduction_dspipelines_challenges}

\paragraph{Missing dependencies specifications} While the notebooks and code were easily accessible, we were missing details about the versions used for the dependencies. To solve this, we inferred the dependencies versions based on the date of release of their reproduction package and compared the results obtained with the ones reported in their study.

\paragraph{Limited parser output} An additional limitation came from the output produced by their parser. Indeed, it is limited to a set of strings (containing acronyms), each one representing a pipeline (\eg~$s = \{A,Pr,M,Pr,A\}$ where \textit{A} is the acronym for \textit{Data Acquisition} for instance). While this may be enough for reporting identified pipelines, it is not possible to link the original code instructions to these stages. Such retrieval is required to properly compare their rule-based classifier with our SLMs approach. We therefore reverse-engineered their code to produce the instruction-to-stage mapping in addition to the original strings. Overall, we were able to extract a dataset, \datasetdspipelines, of $4,975$ instructions mapped to stages from \taxodspipelines.

\subsubsection{Findings}
\label{ss:reproduction_dspipelines_insights}

\paragraph{Partial support of the dataset} Beyond the $4,975$ mapped instructions, we found a total of $16,225$ instructions yielded by the parser, which means that only $30.66\%$ of the code instructions are supported by the mapping. Overall, we identified a relatively large number of distinct function names ($n = 875$), compared to the static mapping size ($n = 404$), and the small number of notebooks analyzed in the reference study ($n = 105$). This highlights the inherent limitation of using a static mapping for this task. Since developers are constantly introducing new functions, manually maintaining such mapping is not sustainable. A more flexible, potentially automated approach is therefore strictly necessary to conduct analyses at a larger scale.

\paragraph{Conflicting duplicated mapping entries} We also analyzed the provided function-to-stage mapping. Among the $404$ entries, we identified $25$ duplicated keys, including $18$ conflicting entries (same function name but different stage mapped). We removed the duplicated non-conflicting entries and fixed the conflicting ones with our experts (also involved in the taxonomy unification in Section~\ref{ss:taxonomies_unification}). Additionally, we reviewed each entry and identified $4$ inconsistencies (\eg~\textit{tqdm} util function labeled as \textit{Modeling}). Overall, this represented $4/379 = 1.05\%$ of the mapping's rules, corresponding to $70/16,225 = 0.43\%$ of all instructions and $70/4,975 = 1.41\%$ of instructions supported by the mapping.

\paragraph{Notebooks are not necessarily ML pipelines} Finally, we reviewed the 105 notebooks and identified that one of them, \textit{class-wise-regex-functions-l-b-0-995.py}, is not an ML pipeline. Indeed, it only contains util functions, and its instructions are only mapped to \textit{Others} in the dataset. We will therefore consider, in the third part of our study (practices exploration), that the insights reported by \textit{DS-Pipelines} correspond to the study of 104 ML pipelines.

\subsection{DASWOW}
\label{ss:reproduction_daswow}

The \textit{DASWOW} dataset,~\datasetdaswow, consists of $470$ notebooks ($9,678$ code cells) organized in $3$ dataframes (cells x metadata): \textit{test\_features} ($1918$x$32$), \textit{train\_features} ($5833$x$32$), \textit{validation\_features} ($1927$x$32$). Each dataframe contains metadata such as the filename, cell\_type, text (code content), etc. For each row, corresponding to a cell of a notebook, columns corresponding to the stage in the taxonomy have the value $0$ or $1$. This taxonomy, \taxodaswow, is different than \taxodspipelines and contains $10$ stages (\cf Table~\ref{tab:daswow_taxonomy} in the Appendix) derived from $4$ studies~\cite{zhang2022coral,garijo2013detecting,wang2022documentation,aggarwal2019can}.
\subsubsection{Challenges}
\label{ss:reproduction_daswow_challenges}

\paragraph{Missing dependencies specifications} Similar to the first reference study, we were missing the dependencies versions. We adopted the same mitigation technique by inferring the versions from release dates and the \textit{Python Package Index} (PyPI).

\paragraph{Missing trained model} More importantly, the trained model was not provided as part of their reproduction package. For that reason, we trained it reusing the provided code and datasets. Overall, the training duration took $2$ hours. The results obtained are essentially identical to the ones reported by \textit{DASWOW} authors, with the same weighted average \textit{F}1 score of $0.71$. Yet, a slight variation of $-0.01$ in the recall, likely due to noise caused by the hardware and potentially different minor versions of the libraries, was observed. We provide the trained model in our reproduction package.

\paragraph{Corrupted dataset's Python code} Because we use the same AST-based parsing methodology for both studies, we needed to access the source code of the notebooks used by~\textit{DASWOW}. However, the code available in their initial reproduction package is corrupted as the indentation is missing. We contacted the authors of the study in September $2025$, who suggested that we obtain the original notebooks directly from the dataset \datasetrule\footnote{\url{https://library.ucsd.edu/dc/object/bb2733859v}} created by Rule et al.~\cite{rule2018exploration}. We therefore retrieved the entire set of $470$ notebooks from the $1.25$ million notebooks dataset. In the meantime, we noticed that the authors of the study had published in November $2025$ the original notebooks in a second supplementary reproduction package\cite{ramasamy_2025_17638925}. This allowed us to compare and validate our collected notebooks. Once validated, we manually fixed some inconsistencies in the code (\eg~bad indentation, Python2 instructions) to finally parse the entire dataset.

\subsubsection{Findings}
\label{ss:reproduction_daswow_insights}

\paragraph{Train and Validation datasets merged} The \textit{DASWOW} dataset consists of $3$ splits: train, test, and validation. However, we noticed that the train and validation datasets were merged by the \textit{DASWOW} authors to train the model. For a fair comparison, and to avoid data contamination, we will only evaluate the model and our selected SLMs on the test split.

\paragraph{Unbalanced classes} As reported by the original study, the dataset is highly unbalanced with $49.5\%$ of the stages being \textit{data\_preprocessing} or \textit{data\_exploration}, and \textit{save\_results} only accounting for $1.2\%$. While this is a good illustration of Data Science in practice, it also highlights the necessity to select appropriate metrics to avoid biased analysis of the classification results.

\subsection{Power Analysis}
\label{ss:sampling_power_analysis}

Before proceeding with our study, we conducted an \textit{a priori} power analysis~\cite{lakens2022sample} to determine if the available dataset size (\ie~minimum between \datasetdspipelines~and \datasetdaswow~sizes) limits the risk for \textit{Type I} and \textit{Type II} errors. We used the \textit{G*Power} software with $\alpha=0.05$ and $1-\beta=0.8$ (power) under the two-tail hypothesis for the non-parametric tests for paired data (McNemar, Cochran's Q) as explained in the literature~\cite{card2020little}. For that purpose, we formulated $3$ scenarios defining the proportion of discordant pairs $\pi\textsubscript{D} = \pi\textsubscript{12} + \pi\textsubscript{21}$ and odds ratio OR $ = \pi\textsubscript{12} / \pi\textsubscript{21}$ (\cf~Table~\ref{tab:power_analysis_sensitivity_mcnemar_cochranq_scenarios}). As a result, the minimum sample size required to conduct our analysis is $796$ (cells for \textit{DASWOW} or instructions for \textit{DS-Pipelines}), which is smaller than the size of our datasets. Same values were used for the goodness-of-fit test as recommended by Berman et al.~\cite{berman2016essential}. We also computed the required sample size for two different degrees of freedom (dof): $1$ and $46$, representing the minimum and maximum dof for our analysis (\cf~Section~\ref{ss:practices_exploration}). Results listed in Table~\ref{tab:power_analysis_sensitivity_goodness_scenarios} indicates a sufficient sample size for all scenarios excepted the conservative with large dof. In other words, we can identify variable effect sizes except for the smallest one with the largest degree of freedom.

\begin{table}
\caption{A priori analysis for McNemar and Cochran's Q tests with 3 scenarios. $\pi\textsubscript{21}$ is the proportion that worsened. $\pi\textsubscript{12}$ is the proportion that improved}
\label{tab:power_analysis_sensitivity_mcnemar_cochranq_scenarios}
\centering
\begin{tabular}{lrrrr}
\toprule
scenario & $\pi\textsubscript{21}$ & $\pi\textsubscript{12}$ & odds ratio & sample size \\
\midrule
Conservative ($5\%$ increase) & 0.10 & 0.15 & 1.5 & \textbf{796}  \\
Moderate ($10\%$ increase) & 0.10 & 0.20 & 2.0 & \textbf{240}  \\
Optimistic ($20\%$ increase) & 0.05 & 0.25 & 1.5 & \textbf{57}  \\
\bottomrule
\end{tabular}
\end{table}

\begin{table}
\caption{A priori analysis for goodness-of-fit tests with 3 scenarios, and two degree of freedom bounds for each scenario}
\label{tab:power_analysis_sensitivity_goodness_scenarios}
\centering
\begin{tabular}{lrrrr}
\toprule
scenario & effect size & dof & sample size \\
\midrule
\multicolumn{3}{l}{\textit{Small dof}} \\
Conservative (small effect) & 0.10 & 1 & \textbf{785}  \\
Moderate (medium effect) & 0.30 & 1 & \textbf{88}  \\
Optimistic (large effect) & 0.50 & 1 & \textbf{32}  \\
\hline
\multicolumn{3}{l}{\textit{Large dof}} \\
Conservative (small effect) & 0.10 & 46 & \textbf{2919}  \\
Moderate (medium effect) & 0.30 & 46 & \textbf{325}  \\
Optimistic (large effect) & 0.50 & 46 & \textbf{117}  \\
\bottomrule
\end{tabular}
\end{table}

However, we found after conducting our experiment that this conservative scenario did not correspond to the results obtained (\cf Section~\ref{ss:practices_exploration}). Indeed, we measured a minimum effect size $\omega = 0.267$, which requires a minimum sample size of $410$. More globally, we found that the other scenarios of our a priori analysis covered the results reported in our experiment (\cf Section~\ref{s:study_results}). We can conclude that all effects reported in our experiment could be detected with adequate statistical power.

%% file: 4_study_design.tex
\section{Study Preparation}
\label{s:study_preparation}

\subsection{Inference Configuration}
\label{ss:inference_configuration}

\paragraph{Structured Output}

We constrained the tested SLM responses to follow a specific JSON schema using ``structured output''\cite{liu2024we}. Thus, the consistency of the LLM output format is optimized which facilitates its automated processing. The schema contains a single field, \textsc{class}, corresponding to the class predicted by the SLM. To further constrain the SLM, we restrict the output to a specific set of ML stages, by typing the \textsc{class} field as an enum. We also needed to parameterize the taxonomy headwords used in the schema, as our study involves different taxonomy versions (mutations). Accordingly, we created the structured output as a Jinja template with ML stages populated based on the experiment configuration. This template is illustrated in Listing~\ref{lst:structured_output}.

\begin{lstlisting}[
    basicstyle=\ttfamily\scriptsize,
    label={lst:structured_output},
    caption={Template for the structured output. Except for field \textsc{class}, all other fields are generic and required by the inference client (OpenAI ``standard''). They are not specific to the templating framework.}
]
{
  "type": "json_schema",
  "json_schema": {
    "name": "classified_line",
    "schema": {
      "type": "object",
      "properties": {
        "class": {
          "enum": {{ compatible_step_names|tojson }},
          "description": "The category."
        }
      },
      "required": [
        "class"
      ]
    }
  }
}
\end{lstlisting}

\paragraph{Hyperparameter Definition}

We controlled SLM-related confounding variables (\cf registered report) by defining a common configuration for all SLMs used in our experiments. This is crucial as inference parameters and prompting techniques can significantly impact classification output~\cite{wei2022chain}. First, we set $\textit{temperature}=0$ and $\textit{top\_p}=1$ (nucleus sampling) to limit predictions randomness. We also set the $frequency\_penalty=0$ and $presence\_penalty=0$ to prevent class variation. Then, the \textit{max\_tokens} parameter was adjusted to $max\_tokens=15$, considering the taxonomy's headwords length and additional tokens related to the structured output schema aforementioned (\eg~curly brackets).

\subsection{SLMs Selection}
\label{s:slms_selection}

\textbf{Answering our research questions involves selecting a pool of SLMs to compare with our reference studies.} This selection process involves multiple criteria.

\paragraph{Models Openness}
To promote openness in SLMs and ensure the reproducibility of our experiments, we exclude closed-weights models (\ie~models that are only accessible through a gated API, with no direct access to their weights). As a results, selected SLMs are either open-source, open-weight or restricted-weight. Restricted-weight models are protected by specific licenses that one must accept before downloading the model's weights. This is the case for popular models such as recent Llama models protected by the Llama 3.2 Community License. These licences often condition their use to a specific purpose (\eg~research only) to avoid any misuse use. More motivation for the inclusion of restricted-weight models is provided in Section~\ref{s:deviations_registered_report}. Among the available SLMs, we focus on those that have been trained for code-related tasks. Being the \textit{de facto} standard, the selected hosting platform for these models is HuggingFace\footnote{\url{https://huggingface.co/}}.

\paragraph{Resources and Reproducibility}

We also consider computing resources and energy consumption as important factors that can limit the reproducibility of our experiments. Indeed, it constitutes a ``Barriers in the culture of research'' as described in the ``Sources of Non-Reproducibility'' by the National Academies of Sciences, Engineering, and Medicine~\cite{national2019reproducibility}. Furthermore, we aim to contribute to greater environmental sustainability, a core principle in ``The Copenhagen Manifesto''~\cite{russo2024generative}, by minimizing the environmental impact of our study. Accordingly, we consider SLMs as language models with a total number of parameters fewer than $8$ billion parameters. Although parameter count is only one factor in energy consumption, such models generally demand lower computational power during inference. Furthermore, we made sure that using these models with appropriate quantization permits reproduction of our study on consumer-grade hardware (tested on a laptop with Nvidia Ada $2000$
GPU and AWQ quantization).

\paragraph{Benchmarks for SLMs Selection}

Following recommendations from~\cite{wagner2025towards}, we selected SLMs based on their performance on the \textit{HumanEval} benchmark~\cite{chen2021evaluating}. Indeed, this benchmark is mature and relevant for code-related tasks. While no official results list is maintained by their creator, some unofficial can be found online. We used the results from \textit{llm-stats}\footnote{\url{https://llm-stats.com/benchmarks/humaneval}}. As of 18 February 2026, the date of selection, the benchmark listed 64 models. Among them, 10 matched our models size and openness constraints. Table~\ref{tab:slm_performance_benchmarks} summarizes the 10 models with their global ranking, number of parameters, and scores.

We can notice a large representation of models from the Qwen provider. While this can be explained by the good performance of their models, they can also be over-represented because of the possible non-exhaustiveness of the hosting platform. For this reason, we applied a filter excluding generalitic Qwen models which did not outperform the coder version: \textit{Qwen2 7B Instruct}, \textit{Qwen2.5 7B Instruct}, and \textit{Qwen2.5-Omni-7B} (multimodal). Separately, we applied a second filter to remove models that were not compatible with our vLLM inference server. This last filter led to the exclusion of \textit{LiteRT} models from Google (specializing in on-device inference): \textit{Gemma 3n E4B Instructed LiteRT (Preview)} and \textit{Gemma 3n E2B Instructed LiteRT (Preview)}. Secondly, we further reduced the potential biases by considering two more recent benchmarks with official leaderboards: \textit{BigCodeBench} \cite{zhuo2024bigcodebench} and \textit{EvalPlus} \cite{evalplus}. Accordingly, we selected the top $3$ SLMs on the $3$ benchmarks to retain the best performing models. This sampling strategy also ensures that each benchmark has the same weight in our evaluation (same amount of models per benchmark). This selection confirmed the presence of Qwen models (\textit{Qwen2.5-Coder-7B-Instruct} and \textit{CodeQwen1.5-7B-Chat}) but also included new providers such as Microsoft (\textit{Phi-3.1-Mini-128K-Instruct}), iSE-UIUC (\textit{Magicoder-S-DS-6.7B}) and Artigenz Labs (\textit{Artigenz-Coder-DS-6.7B}). Overall, we will compare $7$ SLMs, which are highlighted in Table~\ref{tab:slm_performance_benchmarks}.

\begin{table}[!ht]
\caption{Performance of Small Language Models on the HumanEval, BigCodeBench and EvalPlus benchmarks. The selected SLMs for comparison appear in grey. Each SLM can appear in multiple benchmarks.}
\label{tab:slm_performance_benchmarks}
\centering
\renewcommand{\arraystretch}{1.2} % better vertical spacing
\setlength{\tabcolsep}{8pt}       % better horizontal spacing
\begin{tabular}{r l c c}
\hline
\textbf{Rank} & \textbf{Model} & \textbf{\# Params (B)} & \textbf{Score} \\
\hline

\multicolumn{4}{c}{\textbf{HumanEval}} \\
\hline
\rowcolor{lightgray} 19 & Qwen2.5-Coder 7B Instruct & 7    & \textbf{0.884} \\
35 & Qwen2.5 7B Instruct & 7.61 & 0.848 \\
\rowcolor{lightgray} 40 & IBM Granite 4.0 Tiny Preview & 7    & 0.824 \\
44 & Qwen2 7B Instruct & 7.62 & 0.799 \\
45 & Qwen2.5-Omni-7B & 7    & 0.787 \\
47 & Gemma 3n E4B Instructed LiteRT (Preview) & 1.91 & 0.750 \\
\rowcolor{lightgray} 54 & Gemma 3 4B & 4    & 0.713 \\
58 & Gemma 3n E2B Instructed LiteRT (Preview) & 1.91 & 0.665 \\
60 & Phi-3.5-mini-instruct & 3.8  & 0.628 \\
62 & Gemma 3 1B & 1    & 0.415 \\

\hline
\multicolumn{4}{c}{\textbf{BigCodeBench} (average)} \\
\hline
\rowcolor{lightgray} 78 & Phi-3-Mini-128K-Instruct & 3.8 & \textbf{22.0} \\
\rowcolor{lightgray} 86 & Qwen2.5-Coder-7B-Instruct & 7 & 20.3 \\
\rowcolor{lightgray} 102 & CodeQwen1.5-7B-Chat & 7 & 17.2 \\
... & ... & ... & ... \\

\hline
\multicolumn{4}{c}{\textbf{EvalPlus} (average)} \\
\hline
\rowcolor{lightgray} 15 & CodeQwen1.5-7B-Chat & 7 & \textbf{78.850} \\
\rowcolor{lightgray} 21 & Magicoder-S-DS-6.7B & 6.7 & 75.500 \\
\rowcolor{lightgray} 22 & Artigenz-Coder-DS-6.7B & 6.7 & 75.375 \\
... & ... & ... & ... \\

\hline
\end{tabular}
\end{table}

\subsection{Taxonomies Unification}
\label{ss:taxonomies_unification}

\textbf{The reference studies use different taxonomies. In this section, we unify these taxonomies to allow consistent comparisons.} When studying the two reference taxonomies, we noticed that \taxodaswow included elements that are not specific to the ML lifecycle: \textit{helper\_functions} and \textit{comment\_only}. This was confirmed by the original authors definitions (\eg~an helper function is a \textit{``code that is not directly related to the data science activity at hand [...]''}). Because our study specifically targets the ML lifecycle, we decided to remove these two elements from \taxodaswow before proceeding to the taxonomies unification process. To construct the unified taxonomy~\taxounified, we first established a relation $y \subseteq \taxodspipelines \times \taxodaswow$ containing all pairs of equivalent terms between the two reference taxonomies. From this relation, we identified the \textit{connected components} (CC) to group all related terms together. Finally, we defined a mapping function $w$ that assigns a unique \textit{headword} (the unified category name) to each group, such that: $w : CC \to \taxounified$. As shown in Table~\ref{tab:unified_taxonomy}, this process ensures that every term from the original taxonomies is correctly mapped to a single term in our unified framework.

While using a language model for unifying taxonomies could work, it would also introduce uncertainty related to potential hallucinations. Furthermore, we believe that it is important to consult experts whenever possible, especially since this task is not particularly time-consuming. For these reasons, we worked with 3 expert data scientists (\expert{1}, \expert{2}, \expert{3}) from academia, selected for their extensive experience related to Data Science practices. They assisted us in the elaboration of a unified taxonomy~\taxounified. We provided them with the two reference studies to gain a proper understanding of the context. Disagreements were resolved after discussions between the authors and the experts. Table~\ref{tab:taxonomies_unification_kappa} details the overall inter-annotator agreement measured using Fleiss' $\kappa$, and a pairwise analysis using Cohen's $\kappa$. According to~\cite{landis1977measurement}, the overall agreement is almost perfect ($\kappa = 0.89$), as expected given the relative simplicity of the task. In more detail, we can notice that \expert{1} and \expert{3} have a perfect agreement. The only disagreement, between \expert{2} and the others experts, involves the association of ``\textit{Modeling}'' (\textit{DASWOW}) to ``\textit{Training}'' (\textit{DS-Pipelines}). \expert{2} made this distinction because modeling refers to the structure of the algorithm, while training is more related to the training loop. After discussion and consideration of the definitions provided in the reference studies, the experts agreed to merge \textit{Modeling} and \textit{Training}. However, such a distinction could be made in further studies by refining the taxonomy for a more specific need (\cf~Section~\ref{ss:external_validity}).

\begin{table}[ht]
\caption{Inter-annotator agreement measures across the 3 experts.}
\label{tab:taxonomies_unification_kappa}
\centering
\begin{tabular}{l c}
\hline
Metric & $\kappa$ \\
\hline
\multicolumn{2}{l}{\textit{Overall Agreement}} \\
Fleiss' $\kappa$ (3 experts) & 0.89 \\
\hline
\multicolumn{2}{l}{\textit{Pairwise Agreement}} \\
Cohen's $\kappa$ (\expert{1}, \expert{2}) & 0.84 \\
Cohen's $\kappa$ (\expert{1}, \expert{3}) & 1.00 \\
Cohen's $\kappa$ (\expert{2}, \expert{3}) & 0.84 \\
\hline
\end{tabular}
\end{table}

The aforementioned function $w$ is implemented by prompting the best-performing model on HumanEval (\textit{i.e.,} \textit{Qwen2.5-Coder 7B Instruct} as detailed in Table~\ref{tab:slm_performance_benchmarks}) to generate a headword for each element in the relation $y$. Related threats to validity are discussed in Section~\ref{s:threats_validity}. The resulting taxonomy used in our study, \taxounified, is presented in Table~\ref{tab:unified_taxonomy}.

\begin{table}
\centering
\vspace*{\fill}
\caption{Unified taxonomy \taxounified derived from the reference taxonomies}
\label{tab:unified_taxonomy}
\begin{adjustbox}{max width=\textwidth}
\begin{tabular}{lllp{5cm}}
\toprule
unified & DS-Pipelines & DASWOW & definition \\
\midrule
Data Collection & Data Acquisition & load\_data & The process of gathering and importing data from various sources into a data analysis environment for further processing and analysis. \\
Data Preparation & Data Preparation & \makecell[l]{data\_preprocessing,\\data\_exploration} & The initial phase of data science where raw data is cleaned, filtered, transformed, and normalized to ensure quality and relevance for analysis. \\
Data Modeling & \makecell[l]{Modeling,\\Training} & modelling & The process of creating mathematical or computational representations of real-world phenomena using statistical and machine learning techniques to extract insights and make predictions. \\
Model Deployment & Prediction & prediction & The application of a previously trained machine learning model to new, unseen data for the purpose of making predictions or decisions. \\
Model Evaluation & Evaluation & \makecell[l]{evaluation,\\result\_visualization} & The process of assessing a trained model's performance using specific metrics and comparing it against new datasets or real-world scenarios to determine its effectiveness and reliability. \\
Save Results & - & save\_results & The process of serializing and persisting data for future use or analysis. \\
\bottomrule
\end{tabular}
\end{adjustbox}
\vspace*{\fill}
\end{table}

While this taxonomy enables efficient comparison with both reference studies, it comes at the cost of reduced granularity. Indeed, multiple steps from one taxonomy may correspond to a single one in another. First, the distinction between \textit{Modeling} and \textit{Training} is lost. While \textit{DASWOW} does not make this distinction, \textit{DS-Pipelines} reported that these steps were present in similar proportions and often one after the other (\cf~figures $7$ and $8$ in the reference study). We therefore expect that this merging will only have a small effect in the last part of our study. Second, we merged \textit{data\_preprocessing} and \textit{data\_exploration}. While this distinction is made by the authors of \textit{DASWOW}, they often report statistics about these two stages together (\eg~``data exploration and data preprocessing account for $42.9\%$ of the labels''). Finally, we merged \textit{evaluation} and \textit{result visualization}, both related to measurements and assessment of the performance of a model (visually or not). In these 3 cases, further studies could refine and proceed to hierarchical classification to adapt to more specific needs.

\subsection{Prompt Creation}
\label{ss:best_performing_prompt_slm}

\textbf{Before comparing the different SLMs, we need to define a prompt tailored to our classification task. The same prompt structure will be used throughout the course of our study.} We iterated over different prompting techniques as detailed by Schulhoff et al.~\cite{schulhoff2024prompt} to select the best performing one. Overall, we created $9$ main different prompts and evaluated them using the first selected SLM (\textit{Qwen2.5-Coder 7B Instruct}) against the \textit{test features} subset of \datasetdaswow as specified in the registered report. The final prompt version is provided in the appendix (\cf~Listing~\ref{lst:user_prompt}). Given the reported unbalanced stage frequencies in the two reference studies, we evaluated each SLM classification performance using the weighted average \textit{F}1 score and weighted average Matthews Correlation Coefficient (MCC).

\paragraph{Notebook code positioning in the context} We initially introduced the guidance and taxonomy before the notebook code and code instruction to classify. However, we noticed that placing the notebook code in the system prompt as context and moving all other elements (guidance, taxonomy, code instruction) to the user prompt led to an improvement, especially for the weighted average MCC metric, with an increase of $0.024$.

\paragraph{In-Context Learning (ICL) and DASWOW's predictions} We leveraged the ICL ability of language models~\cite{brown2020icl} and provided 8 classification examples. These examples, as well as the number of provided examples, were chosen after multiple iterations and oriented toward classes with poor classification results (\eg~model evaluation). As a result, we observed a $+0.017$ increase for weighted average \textit{F}1 score and $0.014$ for weighted average MCC. Although this technique was the most effective for the \textit{F}1 score metric, the gains remain modest. Inspired by ensemble learning, we also provided the result of DASWOW as a ``special'' example in the prompt (specifying that it could be improved by the model), which further improved the classification results.

\paragraph{Code prompting} We followed Puerto et al. work~\cite{puerto2024code} and rephrased our prompt to a code prompt, to benefit from the coding ability of the language model. Concretely, we wrote a Python function with the taxonomy introduced in its documentation, and each guidance instruction was expressed as a code comment. While this migration alone did not lead to an improvement in performance, it laid the groundwork for more efficient prompting techniques, as detailed below.

\paragraph{Chain-of-Thought (CoT) and taxonomy ordering} Using code prompting allowed us to express the CoT as conditional branches (one per element in our taxonomy). Although the code implementation (\cf function \textsc{classify\_pipeline\_code} in Listing~\ref{lst:user_prompt}) itself was not designed to be functional, its structure helped the model to compare each class. Concretely, we expect the language model to process each element of the taxonomy sequentially rather than as a whole, which could lead to omitted elements. For instance, \textit{Is this code Data Preparation? Otherwise, is it Data Modeling? etc...}. More importantly, this allowed us to evaluate and mitigate potential positional biases\cite{liu2024lost}. Accordingly, we tried different branches ordering (\eg considering Data Modeling before Data Preparation). As a result, we noted a slight improvement compared to the initial taxonomy's elements ordering.

\subsection{Best SLM Selection}
\label{ss:slms_performance_comparison}

\textbf{Having defined an efficient prompt, we can compare the selected SLMs to get the best performing one. This \slmbest will be used for comparison with the reference studies.} We benchmarked each selected SLM on the \textit{test} split of the \textit{DASWOW} dataset. Accordingly, we first conducted a statistical analysis to measure the significance of differences between each pair of SLM (combination resulting in $21$ pairs). Then, we evaluated each model practical performance with weighted average \textit{F}1 score and weighted average MCC. We also report its perplexity value~\cite{bahl1983maximum}, as the average perplexity of generated tokens, to measure the uncertainty of the model predictions.

\paragraph{Statistical significance evaluation}

The heterogeneity in the results was statistically evaluated using the Cochran's Q test with predictions converted to a binary outcome. The Cochran’s Q test revealed a highly significant difference in classification performance among the selected SLMs, $Q = 2906.94$, $p < 0.001$. Consequently, we conducted post-hoc pairwise comparisons with McNemar tests and Holm-Bonferroni correction~\cite{holm1979simple} to determine the specific differences between the SLMs. Overall, we found statistically significant differences in all $21$ pairs, with $\chi^2$ ranging from $11.8518$ to $823.9801$, $p < 0.001$. The maximum $OR = 1/0.049 = 20.38$, obtained when comparing \textit{Qwen2.5-Coder-7B-Instruct} with \textit{Phi-3-mini-128k-instruct}, indicates a large effect. Concretely, \textit{Qwen2.5-Coder-7B-Instruct} reclassifies cases correctly $20$ times more often than incorrectly, compared to \textit{Phi-3-mini-128k-instruct}. The complete results are provided in the appendix (\cf~Table~\ref{tab:all_slms_comparisons_significance}).

\paragraph{Practical evaluation}

\begin{figure}[t]
\caption{\textit{F}1 score vs perplexity for the selected SLMs. The dashed lines represent the median for the \textit{F}1 score (horizontal) and for the perplexity (vertical).}
\label{fig:perf_vs_perplexity}
\includegraphics[width=11cm]{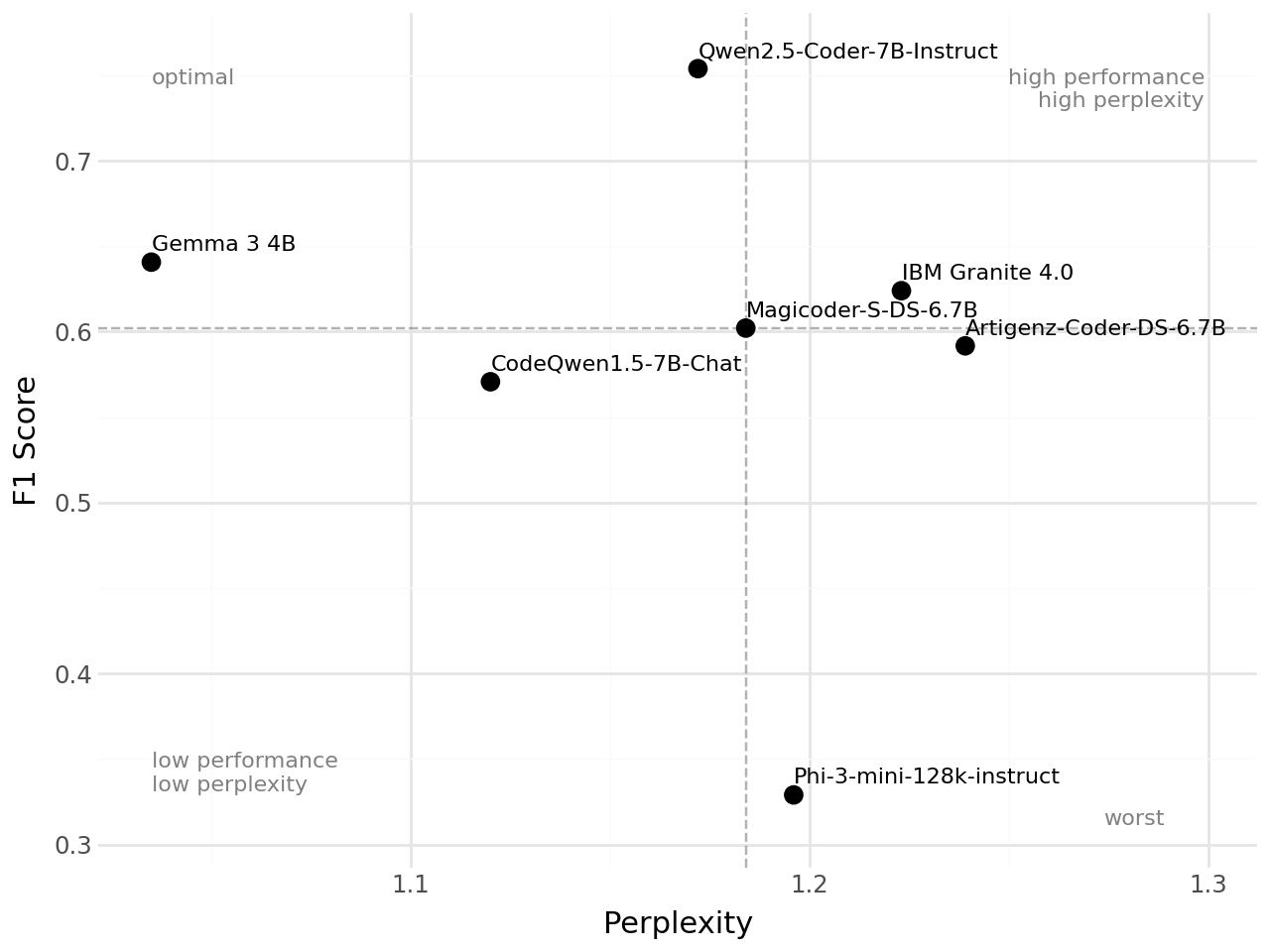}
\centering
\end{figure}

Evaluation results presented in Table~\ref{tab:classif_perf_selected_slms} demonstrate that \textit{Qwen2.5-Coder 7B Instruct} significantly surpasses all other SLMs across the three critical performance metrics, with a notable weighted average \textit{F}1 score of $0.754$. This result implies a substantial advantage over the second-best model (\textit{Gemma 3 4B}), which achieves a weighted average \textit{F}1 score of $0.641$ (margin of $+0.113$), and over the mean baseline performance $0.587$ (margin of $+0.167$). This higher performance, particularly the $+0.425$ gap relative to \textit{Phi-3-mini-128k-instruct}, suggests that \textit{Qwen2.5-Coder 7B Instruct} is more suited for our experiment with our prompting technique. This is a strong argument for its selection for the remaining of our experiments. Regarding the SLM perplexity value, we found that the most efficient model did not have the lowest perplexity (lower is better). This is illustrated in~Figure~\ref{fig:perf_vs_perplexity},
where \textit{Qwen2.5-Coder 7B Instruct} has a higher perplexity than \textit{Gemma 3 4B} for instance. However, \textit{Qwen2.5-Coder 7B Instruct} remains in the optimal location, above the median \textit{F}1 score and below the median perplexity, with a better \textit{F}1 score than \textit{Gemma 3 4B}.

\paragraph{Disparate error rates}

Finally, we measured the error rate as the number of invalid classification output by the SLM. Such errors can be related to badly structured JSON (despite using structured output) or unknown classes (hallucinations). While error rates remain low for $6/7$ of the SLMs (from $0\%$ to $0.152\%$), \textit{Phi-3-Mini-128K-Instruct} presented a very high error rate of $62.674\%$. After investigating, we found that this model struggled complying with the structured output schema and generated invalid JSON.

\begin{table}
\caption{Classification performance of the selected SLMs, on \textit{DASWOW} test dataset}
\label{tab:classif_perf_selected_slms}
\begin{tabular}{lrrrrr}
\toprule
SLM name & accuracy & \textit{F}1 score & mcc & perplexity & error rate (\%) \\
\midrule
Qwen2.5-Coder-7B-Instruct & \textbf{0.586} & \textbf{0.754} & \textbf{0.494} & 1.172 & 0.0095  \\
IBM Granite 4.0 & 0.326 & 0.624 & 0.306 & 1.223 & 0.152 \\
Gemma 3 4B & 0.330 & 0.641 & 0.365 & \textbf{1.035} & \textbf{0} \\
Phi-3-mini-128k-instruct & 0.114 & 0.329 & 0.159 & 1.196 & 62.674 \\
CodeQwen1.5-7B-Chat & 0.507 & 0.571 & 0.225 & 1.120 & \textbf{0} \\
Magicoder-S-DS-6.7B & 0.516 & 0.602 & 0.251 & 1.184 & 0.114 \\
Artigenz-Coder-DS-6.7B & 0.457 & 0.592 & 0.251 & 1.239 & 0.114 \\
\bottomrule
\end{tabular}
\end{table}

%% file: 5_study_results.tex
\section{Study Results}
\label{s:study_results}

\subsection{RQ2. Impact of the Taxonomy Mutations}
\label{ss:taxonomy_mutation}

\begin{markdownquote}
\textbf{H\textsubscript{0.2}} - The choice of words does \textbf{not} significantly impact the classification results.
\end{markdownquote}

\textbf{We have selected the best performing SLM, that we call \slmbest, with an efficient prompt. However, we still need to make sure the taxonomy wording is optimal for \slmbest before comparing the model with the reference studies.} For this reason, we answer \textbf{RQ2} before other research questions. The taxonomy mutation evaluates how robust the SLM's classification performance is to linguistic variations in the stages definitions, a type of perturbation referred to by Chen et al. as ``synonym substitution''~\cite{chen2026nlperturbator}. We evaluate the SLM performance with weighted average \textit{F}1 score and weighted average MCC measures.

\paragraph{Taxonomy Mutation Process and Synonyms Selection}

\begin{figure}[t]
\caption{Example of taxonomy mutation, replacing \textit{Model Deployment} with \textit{Prediction}}
\label{fig:diff_taxonomy_mutation}
\includegraphics[width=11cm]{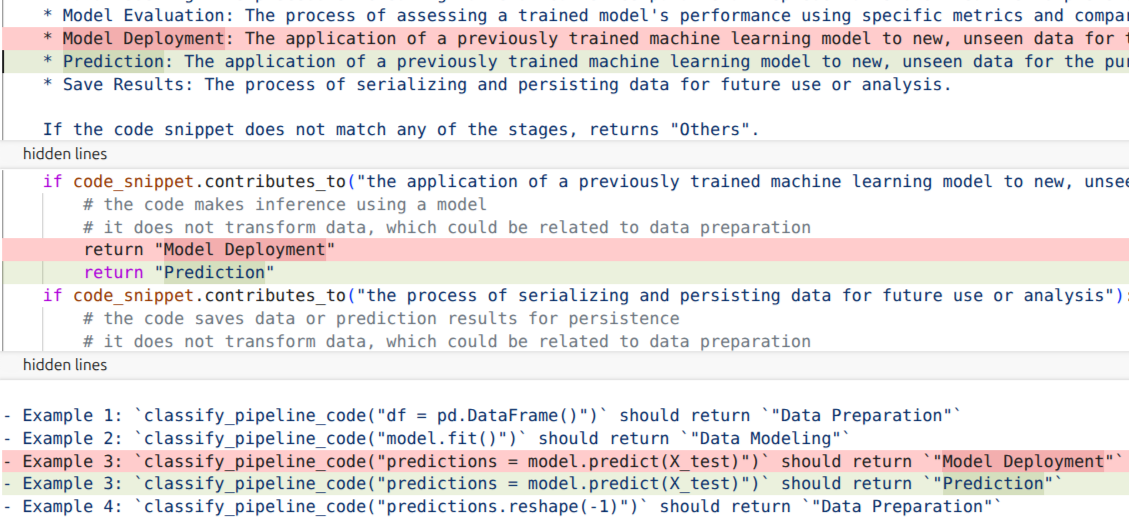}
\centering
\end{figure}

The mutation process consists of creating different versions of \taxounified by replacing a headword with a synonym. Concretely, each occurrence of the headword in the prompt is replaced with its synonym, leaving the associated definitions and comments untouched (\cf~Section \ref{fig:diff_taxonomy_mutation}).
We selected the synonyms from the Data Science domain, rather than standard dictionaries for a more consistent analysis. Specifically, we first collected synonyms used in the two reference studies taxonomies, and asked our experts to provide additional synonyms. As a result, we collected $3$ synonyms per taxonomy element, described in Table~\ref{table:headerwords_synonyms}. We conducted this mutation process on \datasetdspipelines as it is smaller than \datasetdaswow (\ie~requires less resources), and matches the required sample size (\cf Section~\ref{ss:sampling_power_analysis}).

\begin{table}
\caption{Collected synonyms}
\label{table:headerwords_synonyms}
\begin{tabular}{ll}
\toprule
unified & synonyms \\
\midrule
Data Collection & Data Acquisition, Data Gathering, Load Data \\
Data Preparation & Data Preprocessing, Data Wrangling, Data Cleaning \\
Data Modeling & Modeling, Model Building, Model Construction \\
Model Deployment & Prediction, Model Prediction, Inference \\
Model Evaluation & Evaluation, Model Testing, Performance Assessment \\
Save Results & Persist Results, Store Results, Record Results \\
\bottomrule
\end{tabular}
\end{table}

\subsubsection{Results}
\label{sss:taxonomy_mutation_results}

To reject the null hypothesis \textbf{H\textsubscript{0.2}}, we must show that there is a significant difference in the classification results when using \slmbest on \taxounified and any mutated taxonomies.

\paragraph{Statistical significance evaluation} We performed the Cochran's Q test, similarly to Section~\ref{ss:slms_performance_comparison}, with each processing being \slmbest prompted with a different version of the taxonomy ($1$ original and $18$ mutated versions). The Cochran’s Q test revealed a highly significant difference in classification results among the $18$ taxonomies, $Q = 4112.01$, $p < 0.001$. This result confirms that at least one of these mutations significantly impacted the \slmbest classifications. Consequently, post-hoc pairwise comparisons with McNemar tests and Holm-Bonferroni correction highlighted statistically significant differences in $140/171$ pairs, with p-value and $\chi^2$ respectively ranging from $< 0.001$ to $1$, $0$ to $526.08$ and a maximum OR of $9.3$. Statistical values where no significant difference was observed are provided in the appendix (\cf Table~\ref{tab:mutations_comparisons_significance}). All comparisons are available in the reproduction package.

\paragraph{Practical evaluation}

\begin{figure}[t]
\caption{SLM classification performance for the different mutations. The mutation $-1$ corresponds to the original taxonomy.}
\label{fig:classif_mutation_performance}
\includegraphics[width=11cm]{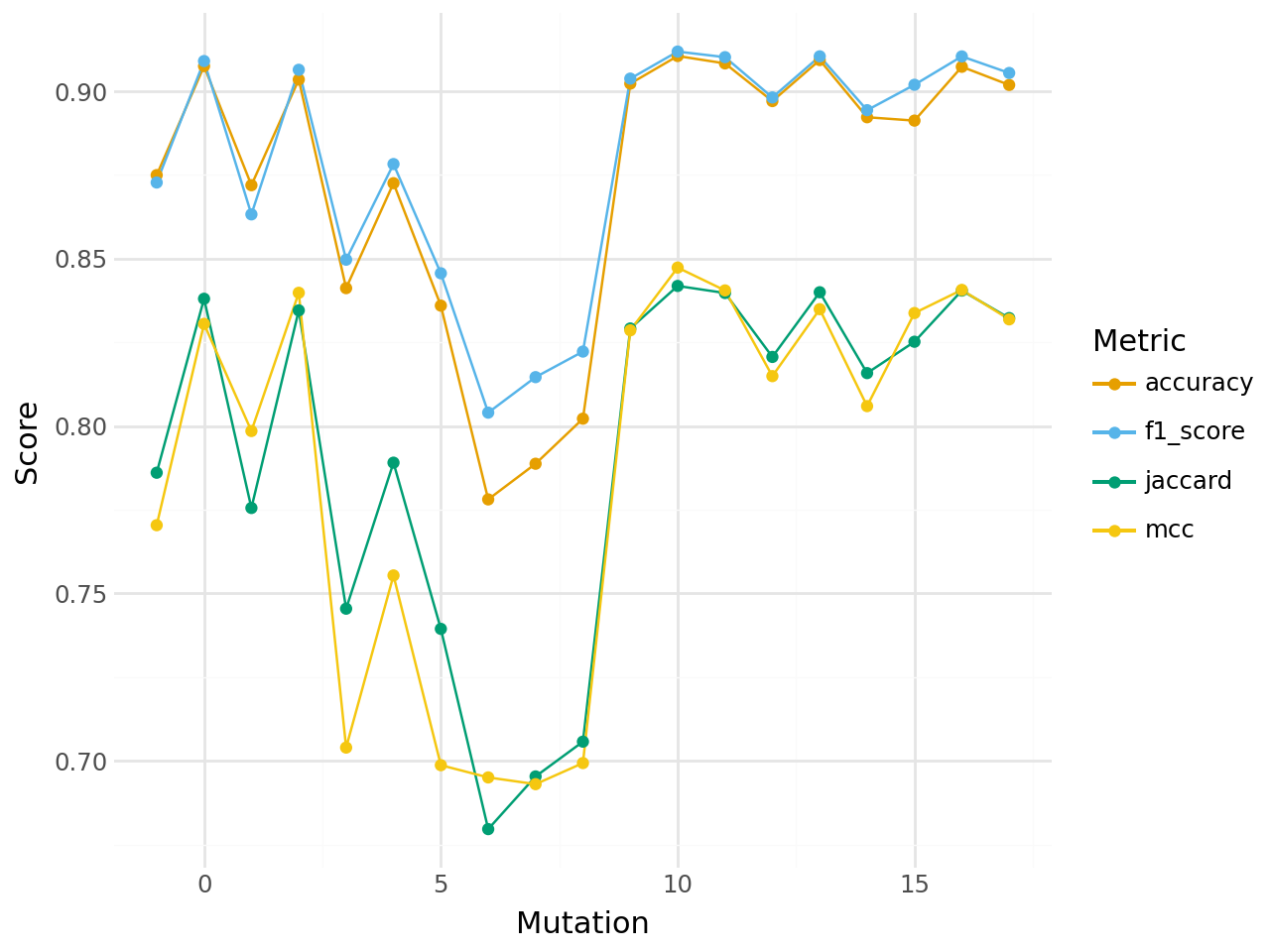}
\centering
\end{figure}

Beyond statistical tests, we also evaluated the practical importance of such mutations using common classification metrics. As illustrated in Figure~\ref{fig:classif_mutation_performance}, we found that some mutations considerably decreased the classification performance, with a maximum difference between two mutations, \textit{Data Modeling} to \textit{Model Building} and \textit{Model Deployment} to \textit{Model Prediction}, of $0.108$ for the weighted average \textit{F}1 score and $0.154$ for the weighted average MCC. We can conclude that the taxonomy wording used in our classification task with \slmbest has both a statistical and practical importance. The best performing taxonomy mutation (mutation $10$) is provided in the Appendix (\cf Table~\ref{tab:best_mutated_unified_taxonomy}).

\begin{tcolorbox}[colback=black!4,colframe=black!50!white]
    \textbf{Finding 1:} The choice of the taxonomy headwords has a statistically significant impact in $82.46\%$ of the mutation pairwise comparisons, with an important effect size (maximum Odds Ratio of $9.3$). It also has a practically significant impact, leading to an increase of up to $+0.108$ for the weighted average \textit{F}1 score and $+0.154$ for the weighted average Matthews Correlation Coefficient (MCC). As a result, \textbf{we can reject the null hypothesis~H\textsubscript{0.2}}.
\end{tcolorbox}

\subsection{RQ1. Performance Comparison between SLM and Reference Studies}
\label{ss:comparison_reference_studies}

\begin{markdownquote}
\textbf{H\textsubscript{0.1}} - SLMs are \textbf{not} more accurate than existing model-based or rule-based classifiers to extract ML pipelines' structures (stages) from ML code.
\end{markdownquote}

\textbf{Having defined a specialized prompt and identified the best performing taxonomy mutation (\cf Table~\ref{tab:best_mutated_unified_taxonomy}), we can now compare \slmbest with the reference studies to answer RQ1.} Evaluating the classification performance of \slmbest against the reference studies allows us to assess the extent to which these language models can serve as an alternative to existing approaches (\textbf{RQ1}). Moreover, results of this comparison will help identify current strengths and weaknesses of SLMs to provide insights about the need for using either fine-tuning or other ML architectures.

\subsubsection{Comparison Process}
\label{sss:comparison_process}

The classification was achieved using \slmbest with the most efficient prompt version and mutated taxonomy. Complying with the registered report, no specific fine-tuning was performed. We extracted code from every notebook of the dataset and parsed it using the AST parser provided by \textit{DS-Pipelines}. Then, we prompted the SLM for each parsed code instruction. Finally, the comparison with reference studies results was made possible after their reproduction and conversion to the unified taxonomy \taxounified using the mapping defined in Section~\ref{ss:taxonomies_unification}.

\paragraph{Comparison with \textit{DS-Pipelines} (Instruction-Level)} Because both \textit{DS-Pipelines} and \slmbest classify code at the instruction level, the comparison was straightforward as we compared each identified instruction's stage with \textit{DS-Pipelines} equivalent.

\paragraph{Comparison with \textit{DASWOW} (Cell-Level and Multi-Label)} The comparison with \textit{DASWOW} was more complex as it classifies each notebook cell (rather than single code instructions) and potentially assigns multiple stages per cell. To enable comparison with our solution, we classified each instruction using \slmbest, group them by respective cell and collected the result set of stages. Then, we compared each cell's set of stages with the \textit{DASWOW} equivalent.

\subsubsection{Results}
\label{sss:reference_studies_comparison_results}

To reject the null hypothesis~\textbf{H\textsubscript{0.1}}, we must demonstrate that \slmbest performs significantly better than existing approaches.

\paragraph{Statistical significance evaluation against DS-Pipelines and DASWOW}

Similar to previous comparisons, we first looked for statistical significance using two distinct McNemar tests, comparing \slmbest to \textit{DASWOW} and \slmbest to \textit{DS-Pipelines}. Both tests revealed statistical significance with respectively $\chi^2(1, N = 1635) = 61.95$, $p < 0.001$, $OR = 0.49$ and $\chi^2(1, N = 4975) = 443$, $p < 0.001$, $OR = 0$. In the case of \textit{DS-Pipelines}, we obtain a null OR value because we cannot improve the predictions of a static mapping. In that case, the objective of our study is to study the gap between our classifier and their static mapping.

\paragraph{Practical evaluation against \textit{DS-Pipelines}}

\begin{figure}[t]
\caption{Normalized mono-label classification matrix of \slmbest classification against \textit{DS-Pipelines}}
\label{fig:dspipelines_confmatrix}
\includegraphics[width=11cm]{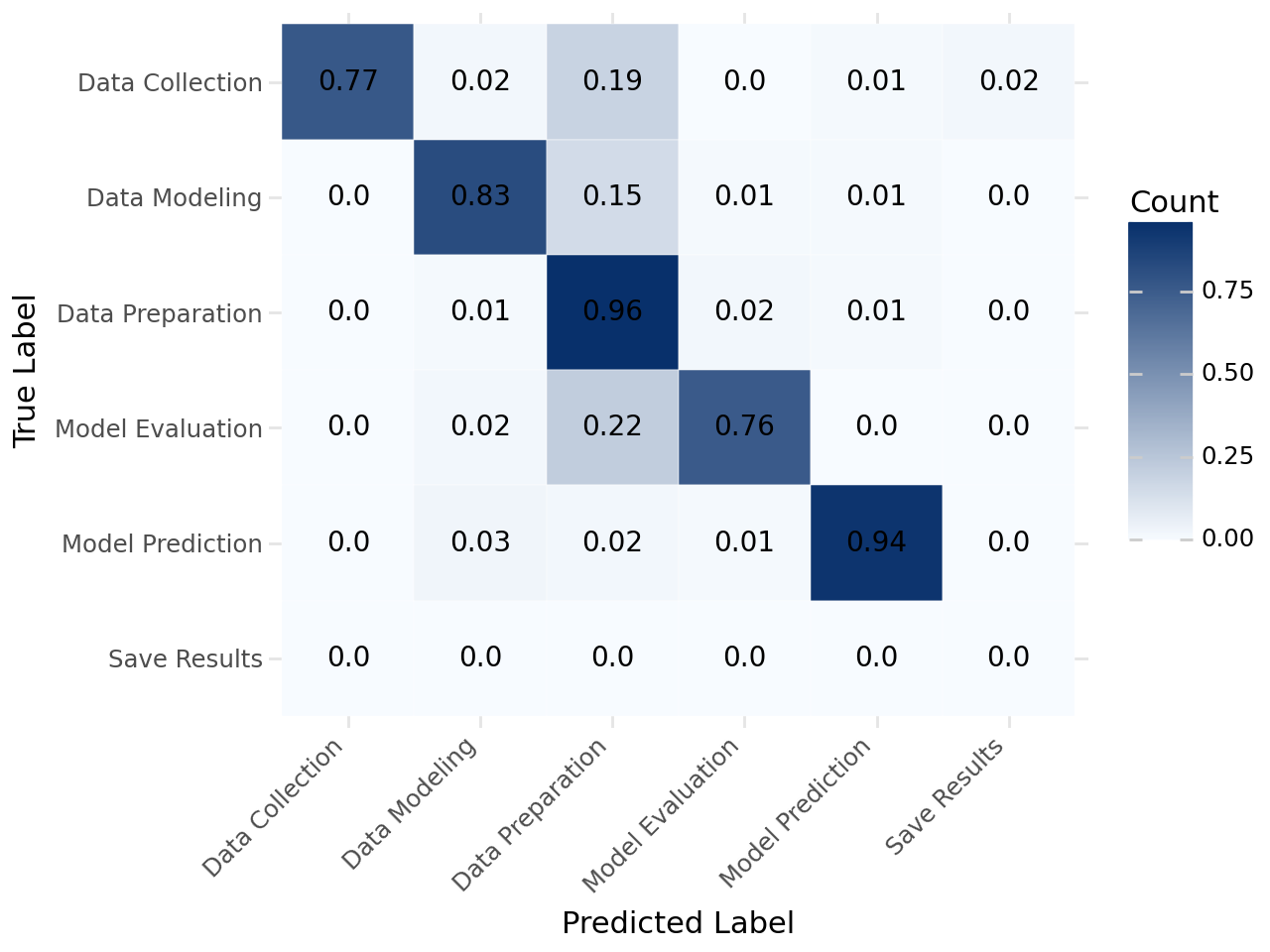}
\centering
\end{figure}

\slmbest achieved a weighted average \textit{F}1 score of $0.912$, subset accuracy of $0.91$ and weighted average MCC of $0.838$. These scores seem to reflect a good ability to classify ML pipeline instructions, without dedicated fine-tuning. However, these scores also hide some disparities in the different classes as illustrated in Figure~\ref{fig:dspipelines_confmatrix}. Indeed, this normalized classification matrix highlights lower performance on \textit{Data Collection} and \textit{Model Evaluation} while describing great results on \textit{Data Preparation} and \textit{Model Prediction}. These disparities are also measured by the MCC score per class, with \textit{Model Prediction} achieving a score of $0.90$, significantly higher than \textit{Model Evaluation} with $0.65$.

\begin{table}[ht]
\centering
\caption{Mono-label classification report for \slmbest on \textit{DS-Pipelines}}
\label{tab:slm_classif_report_dspipelines}
\begin{tabular}{lccccc}
\toprule
\textbf{} & \textbf{Precision} & \textbf{Recall} & \textbf{\textit{F}1 score} & \textbf{Support} & \textbf{MCC} \\
\midrule
Data Collection     & 0.98 & 0.77 & 0.86 & 470  & 0.86 \\
Data Modeling       & 0.95 & 0.83 & 0.89 & 898  & 0.87 \\
Data Preparation    & 0.92 & 0.96 & 0.94 & 3144 & 0.83 \\
Model Evaluation    & 0.58 & 0.76 & 0.66 & 156  & 0.65 \\
Model Prediction    & 0.88 & 0.94 & 0.91 & 307  & 0.90 \\
Save Results        & 0.00 & 0.00 & 0.00 & 0    & 0    \\
\midrule
        &  &  &  &  & \\
Micro Avg      & 0.91 & 0.91 & 0.91 & 4975 & - \\
Macro Avg      & 0.72 & 0.71 & 0.71 & 4975 & 0.68 \\
Weighted Avg   & 0.92 & 0.91 & 0.91 & 4975 & 0.84 \\
Samples Avg    & 0.91 & 0.91 & 0.91 & 4975 & - \\
\bottomrule
\end{tabular}
\end{table}

Finally, we noticed a large representation of \textit{Data Preparation} in the dataset (support in Table~\ref{tab:slm_classif_report_dspipelines}), associated to an over-classification of the same class at the expense of \textit{Data Collection}, \textit{Data Modeling} and \textit{Model Evaluation}. Overall, this represents a high False Positive Rate (FPR) of $14.14\%$, representing $5.2\%$ of the overall dataset. For that reason, the MCC score provides a more nuanced result.

\begin{figure}[t]
\caption{Perplexity per class for \slmbest on \textit{DS-Pipelines}}
\label{fig:dspipelines_perplexity}
\includegraphics[width=11cm]{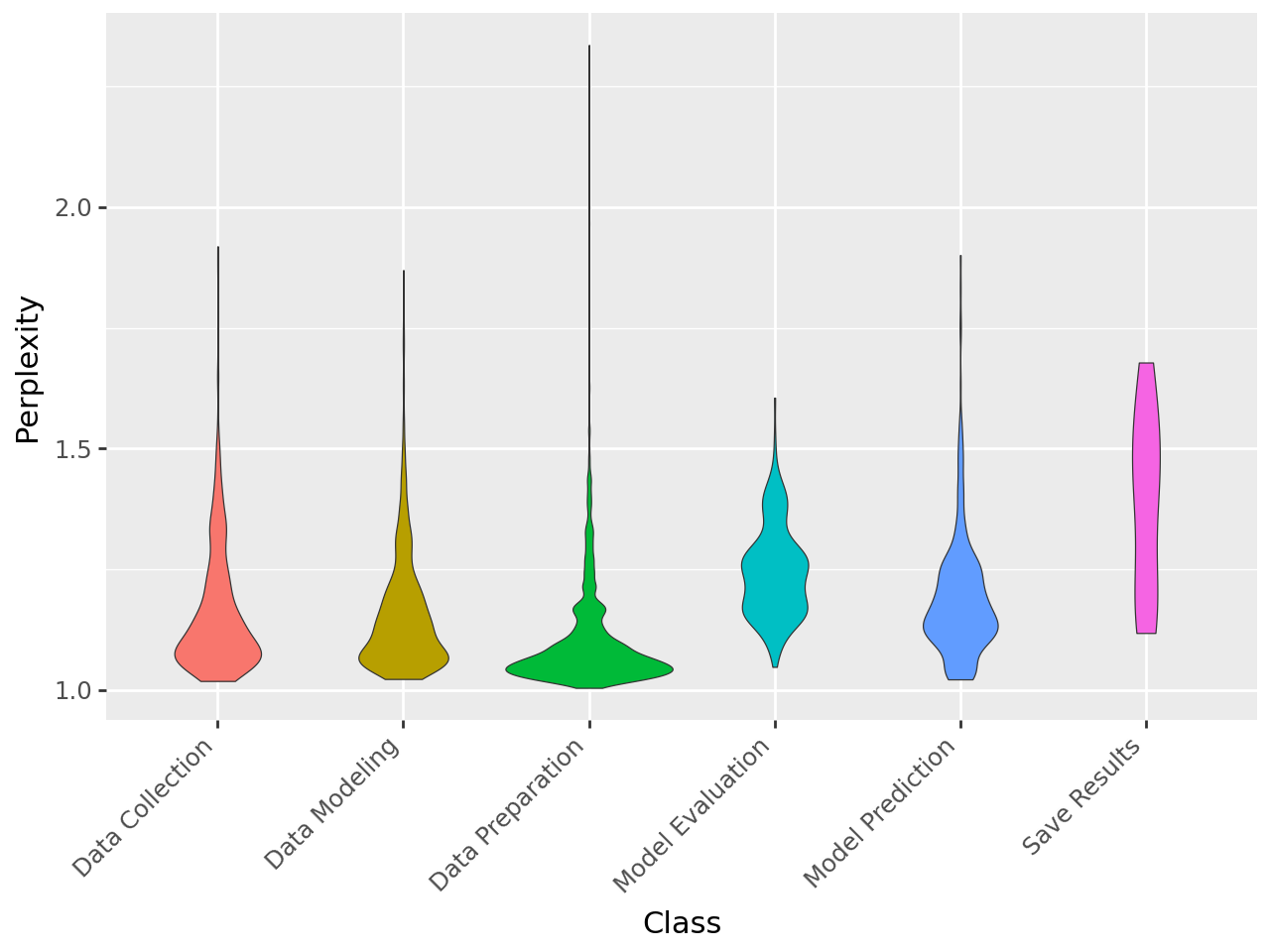}
\centering
\end{figure}

This over-classification is correlated with a very low perplexity (strong confidence) for the same class as represented in Figure~\ref{fig:dspipelines_perplexity}. More globally, we found a relatively low perplexity for all classes (except for \textit{Save Results}), indicating a globally strong ``confidence'' of \slmbest. Beyond this interpretation, it is important to note that perplexity is a metric related to \textbf{tokens} prediction rather than actual class prediction. In the case of \slmbest and mutated \taxounified, a prediction is composed of $2$ tokens: the first word of the headword (\ie~\textit{Data}, \textit{Model}, or \textit{Results}) followed by the second word. In that case, predicting the first token involves choosing among 3 tokens, which contributes to a lower perplexity. Accordingly, $75\%$ of first tokens had a linear probability of being selected $p >= 80.94\%$. We also observed that the second tokens often had an even higher probability, as $75\%$ of second tokens had a linear probability $p >= 94.4\%$. Still, we decided to include all tokens probabilities to compute the perplexity as we also identified some outliers, with respective minimum probabilities of $33.27\%$ and $48.4\%$.

\paragraph{Practical evaluation against \textit{DASWOW}}

The \textit{DASWOW} dataset proved to be a more challenging dataset for \slmbest. While \slmbest achieved a better weighted average \textit{F}1 score in comparison with the \textit{DASWOW} classifier, $0.748$ vs $0.729$, it achieved a significantly lower score on both subset accuracy, $0.545$ vs $0.656$, and weighted average MCC, $0.509$ vs $0.587$. The low results of both classifiers on the subset accuracy is explained by the strictness of this metric. Indeed, it expects predicted labels for a cell to \textit{exactly} match the corresponding set of labels from the true labels (\textit{DASWOW} annotators). For a fair comparison with \textit{DASWOW}, we report the same two complementary metrics, which provide a better picture of the classification performance. First, we calculated the Hamming loss score which measures the fraction of labels that are incorrectly predicted (lower is better). Then, we computed the Jaccard similarity coefficient score which measures the similarity between the sets of predicted labels and the sets of true labels. Overall, \slmbest performed well on the hamming loss score but still fell short compared to the reference classifier, $0.117$ vs $0.088$, and performed similarly in terms of the Jaccard score, $0.609$ vs $0.6$.

\begin{figure}[t]
\caption{MCC per class gaps between \slmbest and the \textit{DASWOW} classifier}
\label{fig:mcc_per_class_gap}
\includegraphics[width=11cm]{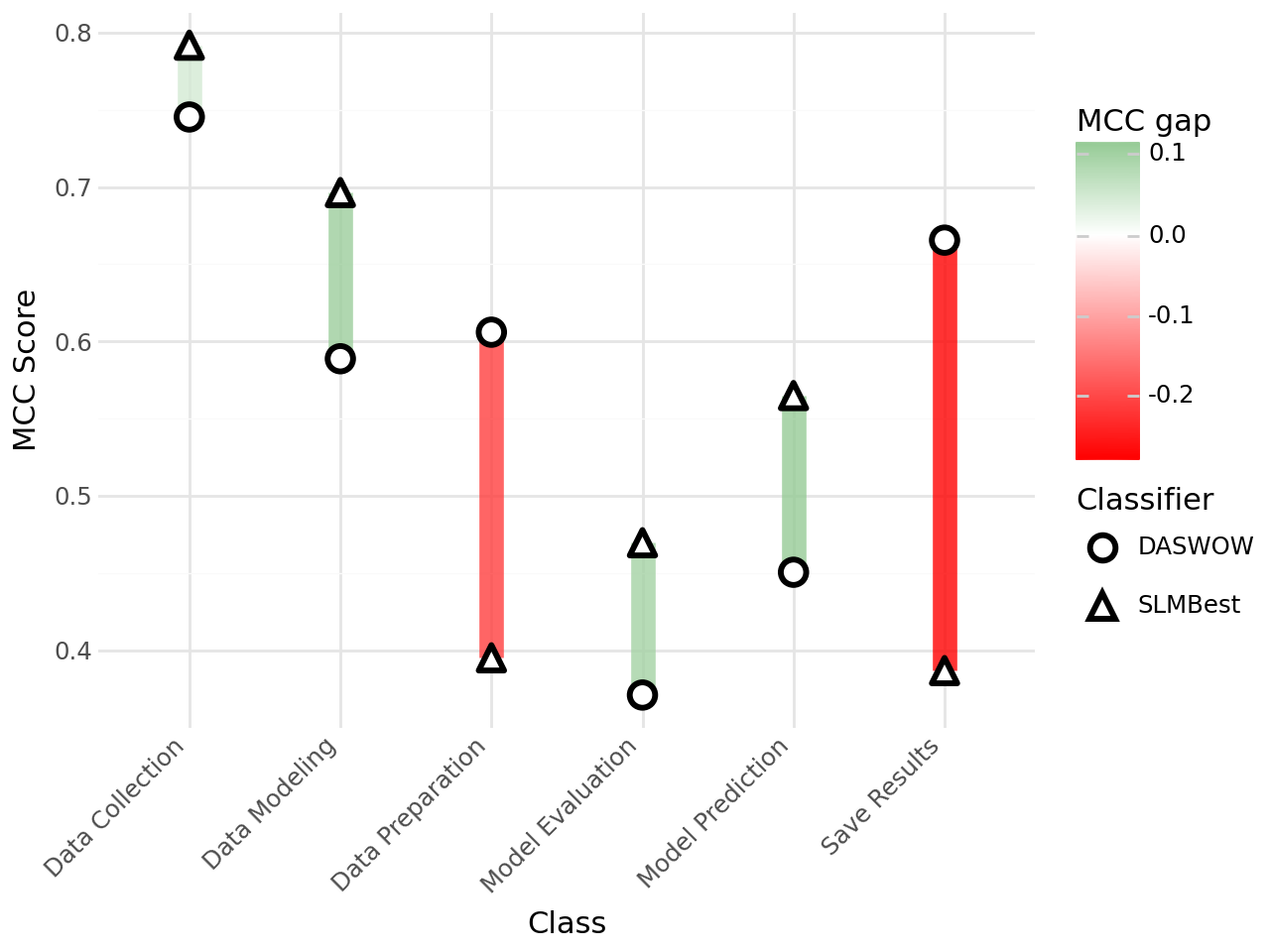}
\centering
\end{figure}

\begin{figure}[t]
\caption{Normalized multi-label classification matrix of \slmbest classification against \textit{DASWOW}}
\label{fig:slm_daswow_confmatrix}
\includegraphics[width=11cm]{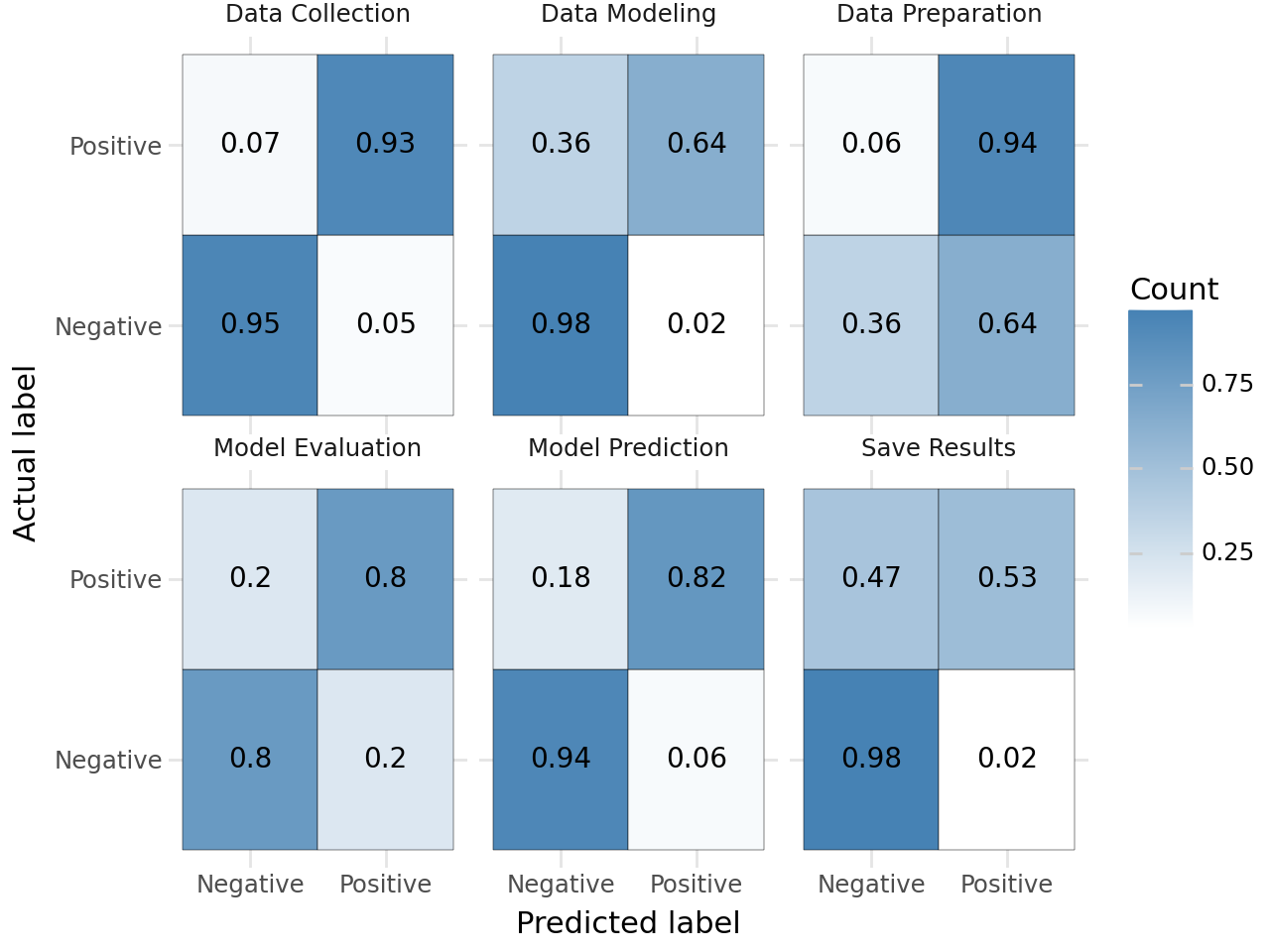}
\centering
\end{figure}

As illustrated in Figure~\ref{fig:slm_daswow_confmatrix}, and similarly to \slmbest on \textit{DS-Pipelines} dataset, there is a tendency for \slmbest to over-classify instructions as~\textit{Data Preparation}. Indeed, $23.55\%$ of the dataset was misclassified as \textit{Data Preparation}, representing a very high FPR of $64\%$. Conversely, \slmbest failed to properly classify a large amount of instructions as \textit{Save Results}, with a false negative rate (FNR) of $47\%$. While this correlates with a low representation of this class in the dataset, this correlation does not extend to \textit{Data Modeling} which accounts for a larger part of the dataset while also yielding a high FNR of $36\%$. In comparison, the \textit{DASWOW} reference classifier had a lower MCC score for all classes, excepted for \textit{Data Preparation} and \textit{Save Results} (\cf~Figure~\ref{fig:mcc_per_class_gap}) which yielded lower FPR, $27\%$, and $0\%$, respectively. Given the high support (weight) of \textit{Data Preparation} ($1030/1887)$, the FPR gap of $37\%$ between the two classification methods can explain the significant difference in the weighted average MCC score of the two classifiers.

\begin{table}[ht]
\centering
\caption{Multi-label classification report for \slmbest on \textit{DASWOW}}
\label{tab:slm_classif_report_daswow}
\begin{tabular}{lccccc}
\toprule
\textbf{} & \textbf{Precision} & \textbf{Recall} & \textbf{\textit{F}1 score} & \textbf{Support} & \textbf{MCC} \\
\midrule
Data Collection     & 0.72 & 0.93 & 0.81 & 187  & 0.79 \\
Data Modeling       & 0.86 & 0.64 & 0.74 & 314  & 0.70 \\
Data Preparation    & 0.72 & 0.94 & 0.81 & 1030 & 0.39 \\
Model Evaluation    & 0.41 & 0.80 & 0.54 & 239  & 0.47 \\
Model Prediction    & 0.43 & 0.82 & 0.57 & 87   & 0.56 \\
Save Results        & 0.30 & 0.53 & 0.39 & 30   & 0.39 \\
\midrule
        &  &  &  &  & \\
Micro Avg      & 0.65 & 0.86 & 0.74 & 1887 & \\
Macro Avg      & 0.57 & 0.78 & 0.64 & 1887 & 0.55 \\
Weighted Avg   & 0.68 & 0.86 & 0.75 & 1887 & 0.51 \\
Samples Avg    & 0.71 & 0.85 & 0.75 & 1887 & \\
\bottomrule
\end{tabular}
\end{table}

\begin{figure}[t]
\caption{Perplexity per class for \slmbest on \textit{DASWOW}}
\label{fig:daswow_perplexity}
\includegraphics[width=11cm]{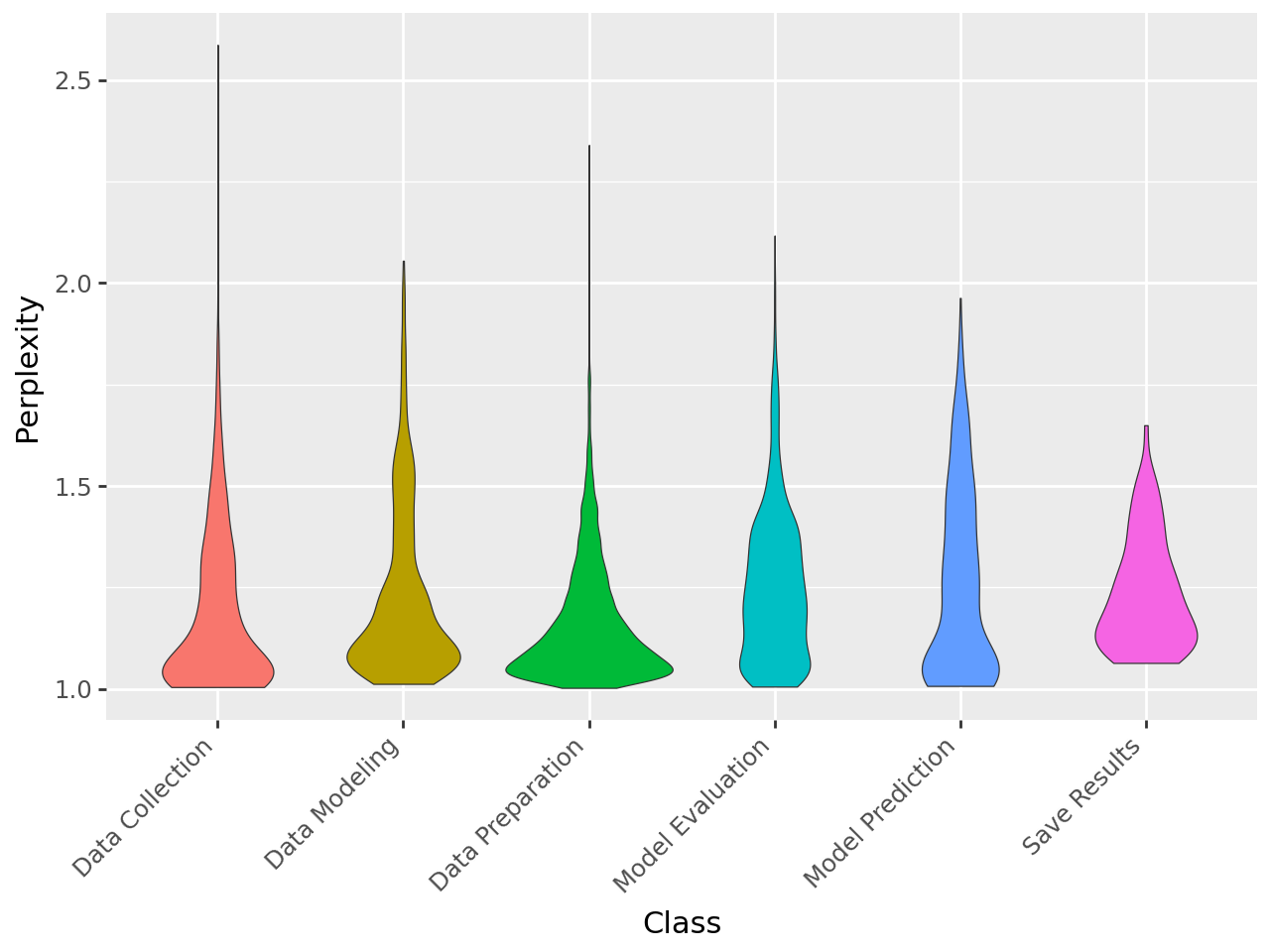}
\centering
\end{figure}

The perplexity obtained on the \textit{DASWOW} dataset is represented in Figure~\ref{fig:daswow_perplexity}. Globally, the average perplexity has increased compared to the one obtained on the \textit{DS-Pipelines} dataset. However, it remains relatively low which indicates that \slmbest is ``confidently'' classifying instructions on both datasets.

\paragraph{Inference duration and energy consumption}

Following recommendations from Henderson et al.~\cite{henderson2020towards}, we report the inference duration and corresponding energy consumption for using \slmbest on our classification task. Our experiments were conducted on the Grid5000 infrastructure~\cite{bolze2006grid} with $1$ Nvidia A100 80GB GPU. The selected inference server was vLLM (v0.10.2) with invariant batching (\textit{VLLM\_\-BATCH\_\-INVARIANT}) to ensure reproducibility of our results. The energy consumption was measured using the \textit{DCGM\_FI\_DEV\_POWER\_USAGE}\footnote{\url{https://docs.nvidia.com/datacenter/dcgm/3.1/dcgm-api/dcgm-api-field-ids.html}} metric, and each \textit{trace} to the inference server was logged and measured using Langfuse (v3.114.0)\footnote{\url{https://langfuse.com/docs}}. Values reported below correspond to the processing of the \textit{DASWOW} dataset, being larger than the \textit{DS-Pipelines} dataset in terms of number of instructions.

Despite having a more limited number of parameters than LLMs, our selected SLMs proved to be very resource-intensive. Indeed, the classification of the entire \textit{DASWOW} test dataset took more than $20$ minutes. At the instruction level, the measured latency (after subtracting the estimated network latency) of $P50 = 0.668s$, $P90 = 2.55s$ and $P99 = 6.77s$ indicates that classifying an instruction within a batch can take up to more than $6$ seconds in $1\%$ of the time. While applying this batch invariance restriction is useful to ensure reproducibility, it does not necessarily represent how such models are used in production. For this reason, we also executed and measured the same experiment with batched queries. Classifying the same dataset took 10:38 minutes. We obtained a better latency for each classified instruction, $P50 = 0.293s$, $P90 = 0.491s$ and $P99 = 1.156s$, at the cost of a non-significant change in the results ($\chi^2 = 0$, $p = 1$, $OR = 1.33$). Nevertheless, the inference duration of \slmbest remains much larger than the reference studies. In comparison, classifying the entire test dataset with the \textit{DASWOW} classifier took approximatively $300$ milliseconds: $P50 = 0.313s$, $P90 = 0.361s$ and $P99 = 0.41s$ (for the entire test dataset).

In terms of energy, we measured a total consumption of $107Wh$ using \slmbest and batch invariance, with a peak at $394W$. On the other side, disabling batch invariance reduced the energy consumption to $29Wh$, with a peak at $318W$. In both configurations, this represents a considerable amount of energy, and raises questions about the suitability of SLMs for larger-scaled analyses.

\begin{tcolorbox}[colback=black!4,colframe=black!50!white]
    \textbf{Finding 2:} Classifying ML code instructions with \slmbest yielded both statistically and practically significant different results than reference studies. Without prior fine-tuning, \slmbest achieved relatively good results on both datasets. However, it did not outperform the reference studies classifiers, especially because of a high False Positive Rate on \textit{Data Preparation}, which has a large support, and strongly impact the weighted average Matthews Correlation Coefficient (MCC). As a result, \textbf{we cannot reject the null hypothesis~H\textsubscript{0.1}}. More work is needed to counterbalance the over-classification of instructions to \textit{Data Preparation}, and mitigate the excessive confidence of the model. Also, the resources required to classify code instructions interrogates the applicability of SLMs for this task at larger scale.
\end{tcolorbox}

\subsection{RQ3. Impact of Classification Methods on our Understanding of ML Practices}
\label{ss:practices_exploration}

\begin{markdownquote}
\textbf{H\textsubscript{0.3}} - Using an SLM does \textbf{not} permit the identification of significantly different practices (patterns) compared to existing studies.
\end{markdownquote}

\textbf{Using \slmbest and the reference studies classifiers, we can investigate the impact of the differences in classification results on our understanding of data scientists' practices.} Reference studies have both provided insights from exploratory analysis of their own datasets. In that context, our goal in this study is not to provide a redundant overview of practices in the field of ML. Instead, we examine the impact of classification methods on our understanding of such practices. In other words, \textit{how reliable are classification results when used to understand the practices of Data Scientists?} To that end, we compare insights from \slmbest classification results with those produced by reference studies' classifiers and the ground truth. Accordingly, the rejection of the null hypothesis~\textbf{H\textsubscript{0.3}} implies the demonstration that using \slmbest leads to the identification of significantly different practices compared to existing studies.

\subsubsection{Exploring the \textit{DASWOW} test dataset}
\label{sss:exploring_daswow_test_dataset}

We analyze the \textit{test} split of the \textit{DASWOW} dataset for two reasons. First, this dataset benefits from a ground truth created by experts in the original study. We use this ground truth to identify the gap between the patterns identified by automated classification and reality. Second, using the \textit{test} split both provides sufficient statistical power (\cf~Table \ref{tab:power_analysis_sensitivity_goodness_scenarios}) and allows comparing with results from the \textit{DASWOW} classifier. Indeed, other splits were used to train this classifier, and using them would have led to data contamination.

\paragraph{Proportion of each stage in the pipelines}

\textbf{Motivation:} Understanding the proportion of the various stages in ML pipelines is a recurrent question in the literature. In particular, it corresponds to \textit{RQ-1a} in \textit{DASWOW}\cite{ramasamy2023workflow} and \textit{RQ4} in \textit{DS-Pipelines}\cite{biswas2022art}. However, it is unclear to what extent the different classification methods provide representative results.

\begin{figure}[t]
\caption{Proportion of stages in ML pipelines. The sum of all proportions per classifier can be lower than $100\%$ (some cells may be unlabelled) or higher than $100\%$ (some cells contain multiple labels).}
\label{fig:proportion_steps_cells}
\includegraphics[width=12cm]{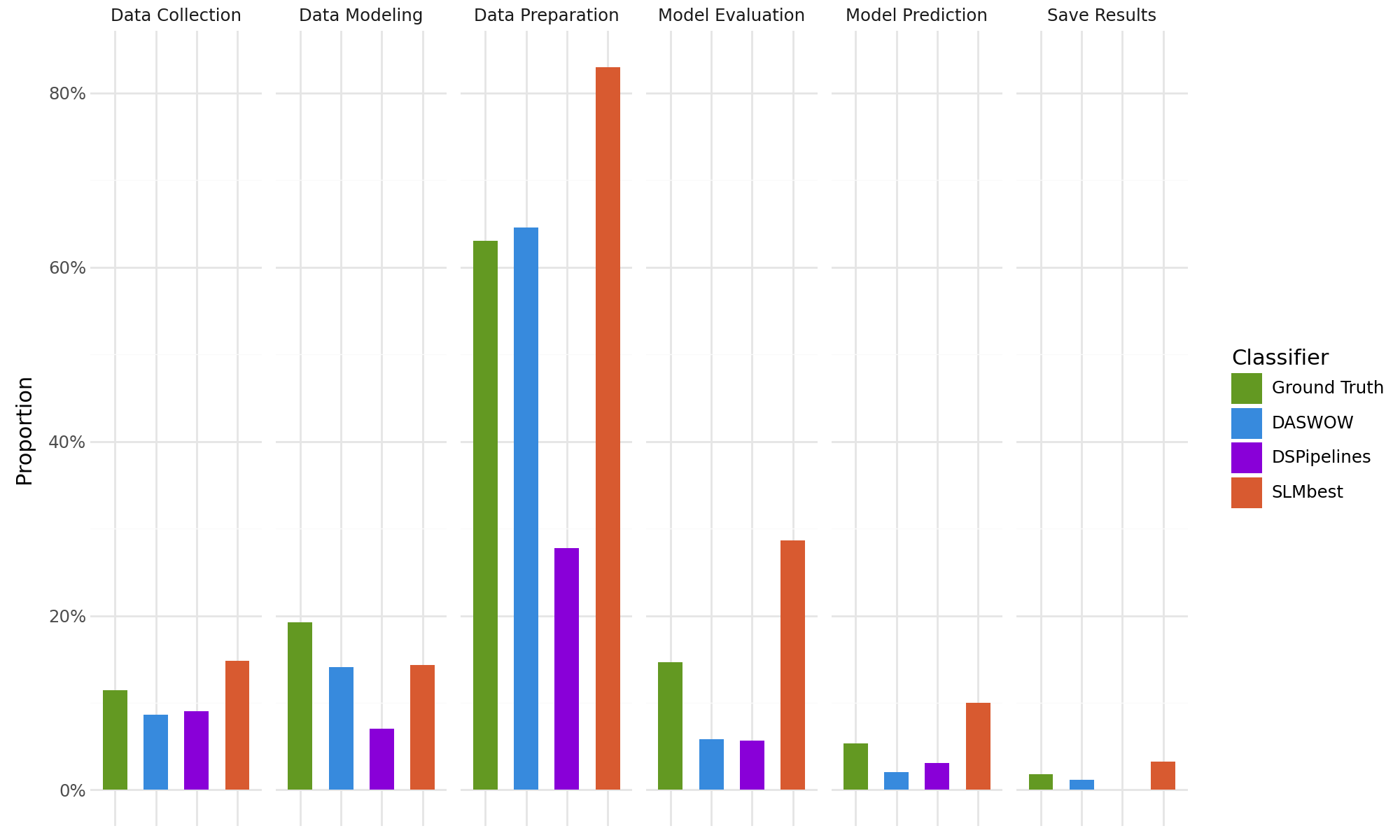}
\centering
\end{figure}

\textbf{Methodology:} Accordingly, we analyzed the proportions of stages present in the pipelines for each classification method and compared them with the ground truth. Figure~\ref{fig:proportion_steps_cells} shows the proportions of stages for each classification.

\textbf{Results:} First, we observe important differences in the reported proportion for \textit{Data Preparation} across classifiers. Indeed, the observed proportion is much larger for \slmbest ($+20\%$ compared to the ground truth), consistent with the over-classification discussed in Section~\ref{sss:reference_studies_comparison_results}. Conversely, classifying with the \textit{DS-Pipelines} static classifier led to an under-representation of \textit{Data Preparation} stages ($-35.23\%$ compared to the ground truth). We attribute this phenomenon to the fact that a large part of \textit{Data Preparation} instructions do not contain any explicit function calls that might appear in the \textit{DS-Pipelines} mapping. Instead, they often consist of arithmetic operations or use Python ``subscript'' (\textit{e.g.,} \textsc{dfLast = df[df['lastRow']==1]}). This underscores the importance of not limiting classification methods to the mapping of function names to stages.

We also observe notable variations for the stages \textit{Model Evaluation} and \textit{Model Prediction}. Again, classifying with \slmbest leads to a substantial over-representation (respectively $+14.01\%$ and $+4.71\%$). As for the \textit{DASWOW} or \textit{DS-Pipelines} classifiers, they under-represent these stages (up to $-8.99\%$ on \textit{Model Evaluation} and $-3.3\%$ on \textit{Model Prediction}). This under-representation (and vice versa) could lead to a misleading interpretation of the dataset, resulting in biased conclusions such as ``Data Scientists infrequently evaluate their models''. More globally, classifying with \textit{DS-Pipelines} systematically leads to an under-representation of the different stages as described by the Mean and Standard Deviation differences ($M = -10.47\%$, $SD = 12.86$) while classifying with \slmbest systematically leads to their over-representation ($M = +6.43\%$, $SD = 9.03$). The \textit{DASWOW} classifier remains a more reliable solution with a lower mean difference and standard deviation ($M = -3.19\%$, $SD = 3.6$). Still, the three classifiers have weaknesses that could lead to misleading interpretations regarding the proportions of the different stages in the studied pipelines.

\begin{tcolorbox}[colback=black!4,colframe=black!50!white]
    \textbf{Finding 3:} Among the evaluated classification methods, \textbf{\slmbest and \textit{DS-Pipelines} led to misrepresented proportions} of the different stages in the dataset, with their respective Mean differences of $M = +6.43\%$ and $M = -10.47\%$. On the other hand, the \textit{DASWOW} classifier appeared to be more reliable, $M = -3.19\%$.
\end{tcolorbox}

\paragraph{Distribution of stages in the pipelines} \textbf{Motivation:} As discussed in the previous paragraph, evaluating the proportion of each stage from automatically classified pipelines remains challenging. However, a classifier may systematically yet uniformly lead to under- or overestimation of proportions. In that case, it would still be possible to answer distribution-related questions such as: ``\textit{How much importance do Data Scientists attach to \textit{Model Evaluation} compared to the other tasks?}''.

\begin{figure}[t]
\caption{Number of occurrences for each classifier compared to the ground truth. The ground truth was scaled to have the same population than the different classifiers. This scaling allows conducting the goodness of fit test.}
\label{fig:scaled_distributions_stages_in_cells}
\includegraphics[width=12cm]{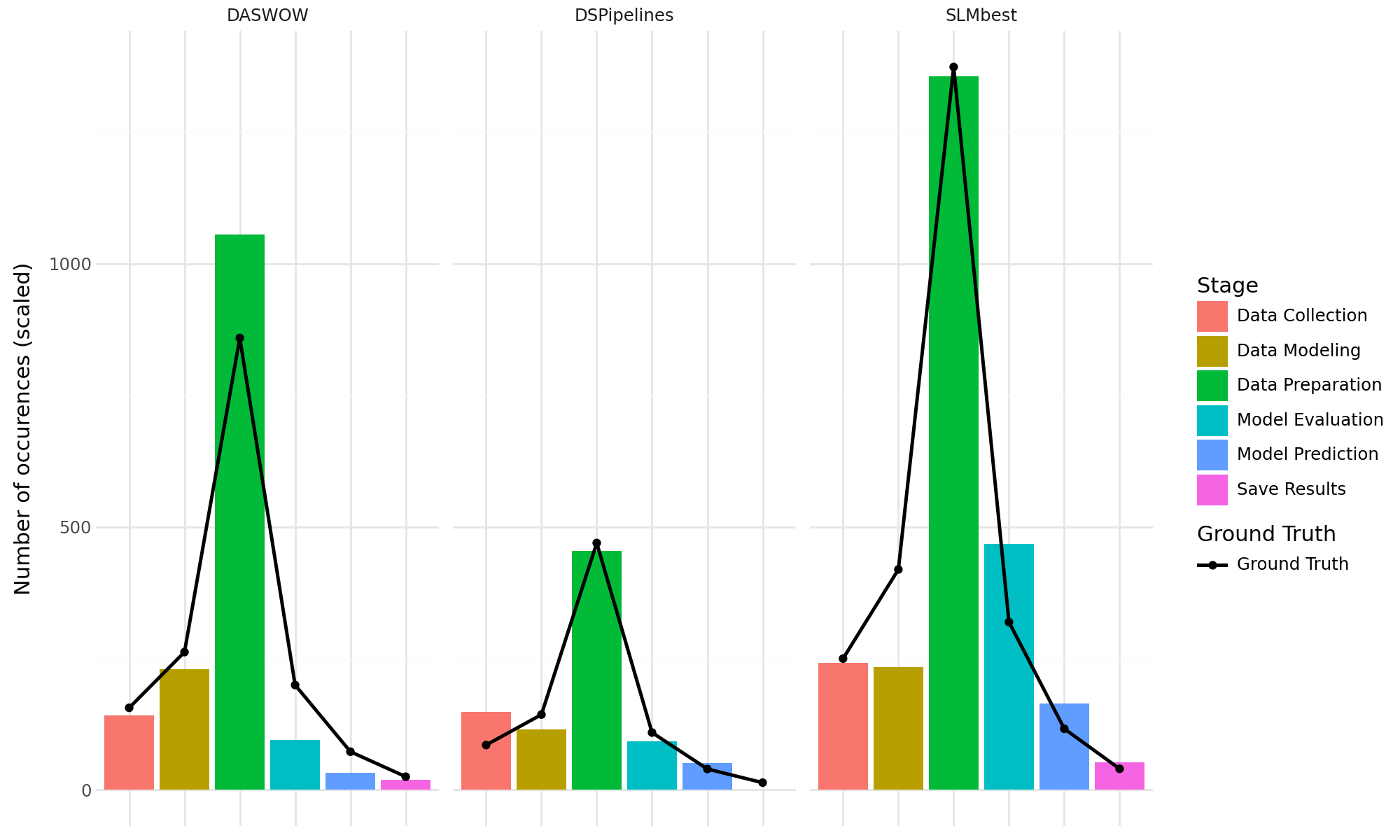}
\centering
\end{figure}

\textbf{Methodology:} Addressing these questions involves quantifying the distribution of the different stages. For this, we conducted one G-test (goodness-of-fit) per classifier, with the ground truth being the expected population and the classifier's results being the observed populations. Because the different population sizes differ (\textit{i.e.,} more or fewer labels per cell compared to the ground truth), we scaled the expected population to match the observed population sizes. The result of this scaling is illustrated in Figure~\ref{fig:scaled_distributions_stages_in_cells}.

\textbf{Results:} Differences between the distributions were found to be significant. \slmbest presented the largest difference, $\chi^2 (5, N = 2518) = 179.95$, $p < 0.001$, followed by \textit{DASWOW}, $\chi^2 (5, N = 1574) = 144.14$, $p < 0.001$, and \textit{DS-Pipelines}, $\chi^2 (5, N = 860) = 77.35$, $p < 0.001$. However, this ranking is also impacted by the number of labels produced by the classifier (the $\chi^2$ test statistic increases with the population size). Accordingly, we measured the effect sizes with Cohen's $\omega$ (small $= 0.10$, medium $= 0.30$, large $= 0.50$)\cite{cohen2013statistical}. We found close small to moderate effect sizes for all classification methods with respectively $\omega = 0.267$, $\omega = 0.303$, and $\omega = 0.300$. This indicates that the observed deviations from the expected distribution are not negligible. Consequently, conclusions drawn from empirical studies that rely on these classification methods should be interpreted with consideration of these noticeable gaps.

The analysis of adjusted Pearson residuals (\cf~Table~\ref{tab:adj_pearson_residuals_distrib_stages} in the Appendix) with Holm–Bonferroni correction revealed that each classifier had significant differences for $50\%$ of the stages. These stages differ across classifiers. Notably, \textit{DASWOW} and \textit{DS-Pipelines} have no significant residuals in common while \slmbest shares $2/3$ of significant residuals with \textit{DASWOW}. Also, the magnitude of this significance ($M = 6.61$, $SD = 2.78$, with absolute values as we consider the two-tailed hypothesis) is very high compared to the thresholds of $2$ or $3$ \cite{agresti2013categorical}. In practical terms, this post-hoc analysis reveals that the observed deviations do not arise from a single stage. Instead, classification weaknesses for half of the stages can lead to misleading empirical conclusions about the distribution of stages across pipelines.

\begin{tcolorbox}[colback=black!4,colframe=black!50!white]
    \textbf{Finding 4:} Differences between the observed stage distribution and the ground truth were found to be \textbf{statistically significant for all classification methods}, with small to moderate effect sizes measured through Cohen's test, ranging from $\omega = 0.267$ to $\omega = 0.303$.
    A post hoc analysis using adjusted Pearson residuals has revealed strong differences attributable to misclassification across multiple stages. Also, the stages playing a key role in this significance differ depending on the classification method used.
\end{tcolorbox}

\paragraph{Transition of stages in the pipelines} \textbf{Motivation:} An ML pipeline is, by definition, a sequence of transitions between %a set of
stages. In that context, understanding the nature and frequencies of these transitions is a prerequisite to the understanding of practices in ML. This topic is also explored in the literature, as illustrated by \textit{DASWOW} and \textit{DS-Pipelines} respective research questions, \textit{RQ-1b} and \textit{RQ5}.

\begin{figure}[t]
\caption{Residual matrices representing the differences between the probability matrices (Markov chains) obtained from the different classifiers and the ground truth. The ground truth was scaled to have the same population than the different classifiers. This scaling allows conducting the goodness of fit test.}
\label{fig:markov_matrices_differences}
\includegraphics[width=12cm]{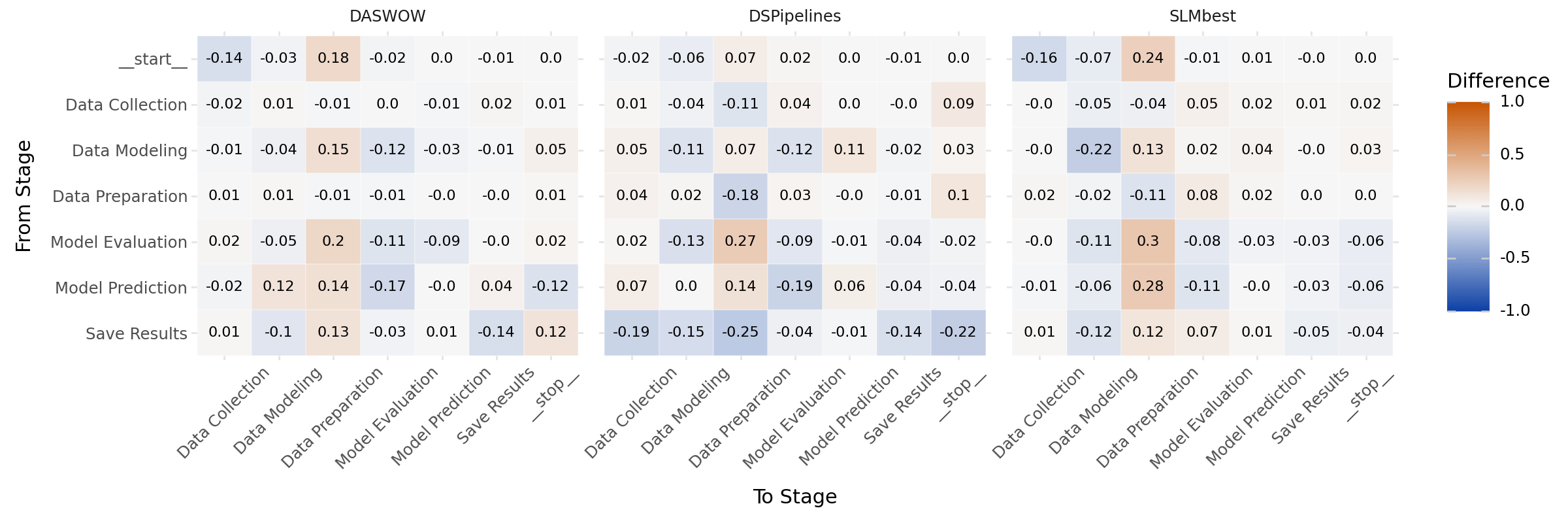}
\centering
\end{figure}

\textbf{Methodology:} We first computed the Markov chains (transition probability matrices) for each classifier. In the case of multiple labels per cell, we followed the methodology of the \textit{DASWOW} study and assigned the probability $p = 1/n$, where $n$ is the number of edges, to each edge in the transition. Then, we computed the residual matrices (the difference between each chain and the expected one from the ground truth) to reveal differences. Once obtained, we flattened each residual matrix and conducted one G-test (goodness-of-fit) per classifier, following the same methodology as the stages distribution analysis. Resulting residual matrices are illustrated in Figure~\ref{fig:markov_matrices_differences}.

\textbf{Results:} The ground truth is composed of $1682$ transitions. Without considering the nature of these transitions, \slmbest was found to be the most accurate with a total of $1728$ transitions ($+2.73\%$). In comparison, \textit{DASWOW} and \textit{DS-Pipelines} yielded respectively $1511$ ($-10.17\%$) and $789$ ($-53.09\%$) transitions. After considering the specific stages present in these transitions, we found statistical differences for all classification methods with varying effect sizes. Similar to our previous analysis of the stages distribution, \slmbest presented the largest difference, $\chi^2 (46, N = 1728) = 584.45$, $p < 0.001$, impacted by the large number of transitions found in the classification results. Interestingly, the \textit{DS-Pipelines} classifier presented the second largest difference, $\chi^2 (46, N = 789) = 372.83$, $p < 0.001$, despite having yielded less transitions than other classifiers. Finally, we found \textit{DASWOW} to be more representative of the transition probabilities from the ground truth, $\chi^2 (46, N = 1511) = 238.76$, $p < 0.001$, despite having a greater difference in the number of yielded transitions compared to \slmbest. Measured effect sizes indicate a moderate effect size for \textit{DASWOW}, $\omega = 0.398$, compared to large effect sizes for \textit{DS-Pipelines} and \slmbest, $\omega = 0.687$ and $\omega = 0.582$ respectively. As a result, \textit{DASWOW} should be preferably used, from a statistical perspective, compared to the other evaluated classifiers.

Yet, a more detailed analysis revealed important discrepancies in the three models that must be considered before conducting further empirical studies. Such discrepancies are first illustrated in the matrix obtained using \textit{DS-Pipelines}, with a systematic under-representation of transitions starting from the \textit{Save Results} stage. More precisely, all transitions from \textit{Save Results} are missed (cumulative probability difference of $-1$) with uneven local differences ($M = -0.14$, $SD = 0.09$). This is explained by the lack of support for the \textit{Save Results} stage in the \textit{DS-Pipelines} mapping (\cf Table~\ref{tab:unified_taxonomy}). Interestingly, these transitions remain infrequent in the ground truth, resulting in the lack of statistical significance of these differences (largest adjusted residual equals to $-1.772$). On the contrary, the difference in the transition from \textit{Model Evaluation} to \textit{Data Preparation} was found to be significant with an important over-representation ($+0.27$) and an adjusted residual value of $4.732$. The same phenomenon also arises with \slmbest and \textit{DASWOW} in different magnitudes, with differences of $+0.3$ and $+0.2$ respectively. As a result, the training of new classification models for identifying stages in ML pipelines should carefully consider this transition by appropriately weighting the training dataset for instance. More globally, the matrix obtained from \slmbest suffers from important differences in transitions to the \textit{Data Preparation} stage ($M = +0.13$, $SD = 0.16$). In practice, this can lead to biased identification of important patterns such as ``\textit{pipeline jungles}''~\cite{2969442.2969519}, where data scientists proceed to \textit{Data Preparation} repetitively between different stages. On the opposite, transitions to the \textit{Data Modeling} are systematically under-represented ($M = -0.09$, $SD = 0.06$). Notably, data scientists construct their models (\textit{Data Modeling}) throughout consecutive notebook cells $+22\%$ more often than the classification results would suggest. In other words, the construction of a model is more sequential than reflected in the classification results. Finally, the residual matrix obtained from \textit{DASWOW} also reveals important discrepancies, despite yielding very accurate transitions from \textit{Data Collection} and \textit{Data Preparation}. For instance, the probability to transition from \textit{Model Prediction} to \textit{Model Evaluation} is under-represented ($-0.17$). Therefore, empirical studies could draw invalid conclusions about the frequency of model evaluation in the pipelines. For instance, wrongfully concluding that ``\textit{data scientists infrequently evaluate their models after having trained them}'' would strongly affect our understanding of practices in ML.

\begin{tcolorbox}[colback=black!4,colframe=black!50!white]
    \textbf{Finding 5:} Differences between the observed stage transitions and the ground truth were found to be \textbf{statistically significant for all classification methods}, with moderate to large effect sizesmeasured through Cohen's test, ranging from $\omega = 0.398$ to $\omega = 0.687$. A more detailed analysis of the practical impact of these differences revealed the potential for biases in empirical studies using the tested classification methods. Based on the previous findings, \textbf{we can reject the null hypothesis~H\textsubscript{0.3}}
\end{tcolorbox}

\subsubsection{Identifying sequential patterns in the \textit{DS-Pipelines} dataset}

In the previous section, we have studied insights from the \textit{DASWOW} dataset (\textit{test} split) at the level of notebook cells. Indeed, a pipeline was represented as a set of notebook cells, each one containing a set of labels. However, understanding the practices in the field of ML implies more fine-grained analyses with the identification of sequential patterns. Such task is impossible with the \textit{DASWOW} classifier as it does not have the required granularity. For instance, it is impossible to know the precise order of stages inside a cell. On the opposite, \slmbest and \textit{DS-Pipelines} work at the level of code instructions which is more precise. Interestingly, we found that \slmbest had a good classification performance on the \textit{DS-Pipelines} dataset, with a weighted \textit{F}1 score of $0.912$. In that context, it is important to understand how the remaining classification errors affect the identification of sequential patterns in the dataset. In this section, we analyze the presence of $3$ sequential patterns in the dataset after classification with \slmbest and the \textit{DS-Pipelines} ground truth. These patterns were selected for their relevance in the literature and their diversity. While the total amount of pipelines remains low ($5,228$ code instructions in a total of $104$ pipelines), the qualitative analysis of potential differences in the arising patterns should provide first indications about the remaining practical gap between \slmbest and \textit{DS-Pipelines}. To that end, we build on a previous case study \cslacroix\cite{lacroixmodel}, adapted to use the mutated unified taxonomy. We extend this case study and consider classification results from \slmbest. Accordingly, we use the \textit{Colombus} platform\footnote{\begin{adjustbox}{max width=0.95\textwidth}{\url{https://archive.softwareheritage.org/swh:1:snp:02fd2031ed8d3dda4cfc9b5d4f9fe8d4987742c9;origin=https://github.com/colombus-dev/colombus}}\end{adjustbox}}, which includes a pattern matching DSL and an interactive interface.

\paragraph{Representative pipeline} Following a manual analysis of their dataset, authors of the \textit{DS-Pipelines} study\cite{biswas2022art} have reported the presence of a recurrent sequence of stages, called ``\textit{representative pipeline}''. In \cslacroix, we have demonstrated that the strict application of this pattern returned no pipeline ($0$ match), while relaxing key constraints (such as making model evaluation optional) matched $54$ pipelines. However, the adaptation to the mutated unified taxonomy led to an increase in the amount of positive matches from $54$ to $64$ for the flexible version. This change is the result of a reduced granularity caused by the taxonomies unification process. More specifically, this change arises from the grouping of \textit{Modeling} and \textit{Training} under a common stage \textit{Data Modeling} in \taxounified (\cf Table~\ref{tab:unified_taxonomy}). Accordingly, we retain values $0$ and $64$ for comparison with \slmbest. We found that using \slmbest for code classification slightly impacted the pipeline structures in the dataset. As a result, the same pattern was found $2$ times in its strict version ($+1.92\%$), and $62$ times in its more flexible version ($-1.92\%$). These differences have little practical impact when considering the overall number of matches. However, we found more variations after computing the symmetric difference between the two sets of matching pipelines obtained from the datasets classified with \textit{DS-Pipelines} or \slmbest. Precisely, we found $8$ different pipelines for the flexible version while we would have expected $2$ (\textit{i.e.,} $64 - 62$). This indicates the presence of compensating errors, in which false positives and false negatives balance each other out. In our case, $3$ expected matching pipelines were replaced with $3$ unexpected ones. This observation further demonstrates the instability of using \slmbest for identifying stages in ML pipelines, and illustrates the large test statistics measured in the other parts of our study.

\paragraph{Feedback loops} Because the construction of ML pipelines is iterative and incremental, stages can be connected with ``\textit{feedback loop}''. For instance, a data scientist can train multiple models, compare their performance, and select the best one. In that case, a feedback loop involves the initial presence of \textit{Data Modeling} stage, followed later in the pipeline by \textit{Model Prediction}, and then a return to \textit{Data Modeling}. Based on this observation, we measured the occurrences of two specific feedback loops represented in the \textit{DS-Pipelines} study: 1) the initial presence of \textit{Data Preparation}, followed later by \textit{Model Prediction} and a return to \textit{Data Preparation}, 2) the initial presence of \textit{Data Modeling}, followed later by \textit{Model Prediction} and a return to \textit{Data Modeling}. In \cslacroix, we found $ 52$ and $8$ matching pipelines, respectively. Again, the adaptation to the mutated unified taxonomy led to an increase in the amount of positive matches from $8$ to $13$ for the last feedback loop. Accordingly, we retain values $52$ and $13$ for comparison with \slmbest. Applying the two patterns on the dataset classified with \slmbest returned $47$ ($-0.96\%$) and $12$ ($-4.81\%$) matches, respectively. Similar to the representative pipeline, the measured matching differences remain low. Additionally, the analysis of symmetric differences confirmed the previous observations, with $6$ compensating errors ($5.77\%$ of the pipelines) present in both feedback loop patterns.

\paragraph{Pipeline jungles} Another interesting pattern arising from ML pipelines consists of preparing data at multiple locations in the pipeline, which can introduce technical debt and limit maintainability of the pipeline. This pattern is called ``\textit{pipeline jungles}''~\cite{2969442.2969519}. Notably, the pattern involves the presence of \textit{Data Preparation} stages, which we found to be challenging for classifying with \slmbest. Looking for the presence of this pattern in the different pipelines, we have identified $65$ matches in the dataset classified with \textit{DS-Pipelines} as opposed to $58$ in the same dataset classified with \slmbest. This represents a decrease ($-6.73\%$) which correlates with the high false positive rate of $14.14\%$ measured in Section~\ref{ss:comparison_reference_studies}. In fact, we found that many of the false positives for \textit{Data Preparation} were located between two \textit{Data Preparation} stages. This artificial merging therefore gives the impression of a less fragmented preparation of the data. Finally, we found $12$ compensating errors which further highlights the need to consider the individual results when exploring the dataset classified with \slmbest.

\begin{tcolorbox}[colback=black!4,colframe=black!50!white]
    \textbf{Finding 6:} Identifying recurring sequential patterns such as feedback loops and pipeline jungles is important to properly characterize the practices of data scientists. While the \textit{DASWOW} classifier does not provide a sufficient granularity, \textbf{classifying with \slmbest only slightly impacted the overall number of patterns found} compared to \textit{DS-Pipelines}. However, a more detailed analysis highlighted the \textbf{lack of stability of \slmbest with the recurrent presence of compensating errors} (up to $11.54\%$).
\end{tcolorbox}

%% file: 6_deviations_registered_report.tex
\section{Deviations From The Registered Report}
\label{s:deviations_registered_report}

Improving the selection process of candidate SLMs is the main deviation from the registered report. We considered two additional benchmarks to reduce potential biases, and better reflect the recent evolution of the practices in the evaluation of language model performance. As a result, we compared 7 SLMs, instead of 5 as initially planned.

Another important variation involves the taxonomy mutation process, where we sourced the list of synonyms from experts of the domain and reference studies papers rather than traditional dictionaries. This avoid biasing our mutation results with synonyms that are not specific to the domain.

We also relaxed the openness selection criteria to include restricted-weights models. This corresponds to an evolving practice in the domain where providers condition the use of language models to specific use case or domains (\eg~research only, prohibited misuse for malicious purposes). However, the only selected restricted-weights model in our study is \textit{Gemma 3 4B}, and no closed-weights model was used.

Finally, we changed the implementation of the goodness of fit tests from Pearson chi-squared to G-tests. This change was necessary because the independence assumption of the Pearson chi-squared test was not met. Indeed, the classification of one cell could be impacted by the nature of the surrounding ones (\textit{e.g.,} \textit{Model Evaluation} implies the presence of \textit{Data Modeling} in the pipeline). However, this change does not affect our evaluation as it remains a goodness-of-fit test. We further reported adjusted Pearson residuals with Holm-Bonferroni correction when it was of interest for the study.

%% file: 7_threats_to_validity.tex
\section{Threats to Validity}
\label{s:threats_validity}

\subsection{Internal Validity}
\label{ss:internal_validity}
Reusing datasets from the two peer-reviewed reference studies may pose a threat as we do not control their diversity (\eg~libraries) and quality. 
However, this choice avoids adding new biases from new annotators and does not prevent us from measuring this diversity. Additionally, we have assessed their quality in collaboration with experts of the domain before conducting the experiments. Another threat comes from the potential contamination with \datasetdspipelines~or \datasetdaswow~being present in the training datasets of selected SLMs. As we do not necessarily have access to the original training datasets, we cannot confidently determine whether contamination has occurred. This is an inherent risks of using language models. Yet, the design of our study has the advantage of using a unique taxonomy, unknown to the model. Furthermore, the risk remains limited as we prompt \slmbest~for each raw line of code while \datasetdaswow~contains the mapping for each cell and \textit{DS-Pipelines} limits the mapping to the function name. It is also possible to have contamination in the different benchmarks used for selecting the SLMs. We mitigate this risk by selecting multiple benchmarks and following recent recommendations~\cite{wagner2025towards}. Also, the prompting technique selection has been done using only one of the SLMs, which may give it an advantage over the other models (\cf Section \ref{ss:slms_performance_comparison}). Testing all prompting techniques on all selected SLMs does not seem feasible in terms of time and available computing resources. However, we limit this risk by considering the best performing model on robust benchmarks. Lastly, the definition of the unified taxonomy involves $3$ experts from our research laboratory, which may add a bias in the annotation process and resulting taxonomy. However, they were selected for their experience in the ML domain, and provided with taxonomies from the literature. The proper evaluation of their agreement further mitigates this risk.

\subsection{External Validity}
\label{ss:external_validity}

The selection of SLMs specifically trained for code-related tasks seems reasonable as we study ML code. However, further studies are needed to evaluate the interest of such restriction. Furthermore, using language models involve having a knowledge cutoff that can lead to a decrease of performance in the future. This phenomenon should be properly evaluated in the next few years. Additionally, our analysis relies on two studies which may limit the generalizability of our conclusions to the full diversity of ML domain. This is also true for the taxonomy mutation, where significant differences could be observed on one dataset but not on another. However, the selection process involves an evaluation of their relevance, and the detailed methodology, along with a reproduction package, should facilitate complementary empirical studies. The decision to limit reference studies to those offering accessible reproduction packages may prevent us from finding more effective solutions, but it promotes open science and ensures our comparisons are based on reproducible artifacts. Finally, the unified taxonomy \taxounified as introduced in our study is not intended to cover all scopes and specific domains of ML. For instance, training a language model involves more specific stages (\eg~training the tokenizer, aligning the model). Similarly, researchers may study more fine-grained data preparation stages (\eg~feature engineering, data filtering). Supporting such pipelines involves extending the taxonomy and the classifier (with hierarchical classification\cite{silla2011survey} for instance).

\subsection{Construct Validity}
\label{ss:construct_validity}

One threat to construct validity regarding the evaluation of classification performance comes from the unbalanced nature of stages. For instance, the data preparation stage represents a more extensive part of the ML pipeline than model evaluation. We mitigate this threat by using the Matthews Correlation Coefficient (MCC) which supports well unbalanced classes in classification tasks. We supplement this analysis with the \textit{F}1 score for each stage to measure the performance on each stage independently. The broad scope of the ``exploration insights'' construct, encompassing the diverse practices of data scientists during ML pipeline development, also poses a threat. We mitigate this by relying on reported insights from the reference studies as reflective indicators to measure it. However, reporting additional insights could deepen the analysis. For this reason, we also conduct a more qualitative analysis with the identification of sequential patterns to enrich the understanding of the construct.

%% file: 8_conclusion.tex
\section{Conclusion and Future Works}
\label{s:conclusion_future_works}

Using SLMs to reverse-engineer ML pipeline structures has shown promise, although many parameters can affect the classification performance, and some limitations have been identified.

First, we found that the choice of the SLM, both statistically and practically, has a significant impact on the classification results. Specifically, \textit{Qwen2.5-Coder 7B Instruct} (\slmbest) outperformed other selected candidates, with a maximum odds ratio of $20.38$.

Second, we found that mutating the taxonomy wording with synonym substitution led to statistically significant differences in $82.46\%$ of the mutations. From a practical perspective, choosing the most appropriate headwords led to important increases of $+0.108$ in the \textit{F}1 score and $+0.154$ in the MCC.

Third, we found that \slmbest~demonstrated good classification performance, reaching \textit{F}1 scores of $0.912$ and $0.748$ on the \textit{DS-Pipelines} and \textit{DASWOW} datasets, respectively. Although it does not outperform the reference studies, it proves competitive with their classifiers while exhibiting statistically significant differences from them. Measured perplexity showed that \slmbest~was highly ``confident'' in its results, while not always corresponding to the expected ones. For instance, \slmbest~over-classified instructions as \textit{Data Preparation}, with false positives representing up to $23.55\%$ of the dataset. Complementary measures highlighted a significantly longer inference duration compared to reference studies, with a total duration of $20$ minutes on the \textit{DASWOW} test dataset. This, along with a high energy consumption of $107Wh$ to classify the same dataset, questions the relevance of using such models in our use case and highlights potential scaling limitations.

Fourth, we found that the choice of the classifier significantly affects our understanding of data scientists' practices. In our experiment, classifying with either \slmbest~or reference classifiers has yielded patterns different from those obtained by human labeling.

Overall, reported results highlight the importance of defining an efficient classification method to properly analyse and understand the practices of data scientists. Future works should go beyond static mapping, which cannot cover the large domain of data science ($27.22\%$ of code instructions were supported by \textit{DS-Pipelines} mapping on its own dataset). The solution must also address cost considerations and remain scalable. Accordingly, we plan to investigate the use of more traditional supervised classifiers (\textit{e.g.,} SVM using active learning) with LLM embeddings, which should mitigate the cost of classifying with language models while benefiting from their extensive training dataset.

%% file: 9_declarations.tex
\begin{acknowledgements}
We would like to thank Anne Etien and Marius Mignard for their valuable feedbacks. We are also grateful to Diane Lingrand and Mathieu Lacage who contributed to the taxonomies unification process.
\end{acknowledgements}

\noindent\textbf{Author Contributions} Material preparation, data collection and analysis were performed by Nicolas Lacroix. The first draft of the manuscript was written by Nicolas Lacroix. Mireille Blay-Fornarino, Frédéric Precioso and Sébastien Mosser provided reviews, feedback, and edits on the manuscript. All authors read and approved the final manuscript. \\

\noindent\textbf{Corresponding Author} Correspondence to \href{mailto:nicolas.lacroix@univ-cotedazur.fr}{Nicolas Lacroix}.

\section*{Ethics Declarations}
\label{s:ethics_declarations}

\noindent\textbf{Funding} This work is funded by the FATES-MLOps international cooperation project (ANR (TSIA 2024) ANR-24-IAS2-0002-01, NSERC – ALLRP-2024-599951) and by the French National Research Agency (ANR) under grant agreement no. ANR-24-CE25-1286-01. \\

\noindent\textbf{Conflict of Interest} The authors declare that they have no conflict of interest. \\

\noindent\textbf{Ethical Approval} Not applicable. This study did not involve any experiments on human subjects. The only experts involved in this study are members of our research laboratory who helped establish the unified taxonomy. \\

\noindent\textbf{Informed Consent} All experts provided informed consent before taking part in the taxonomy unification process. \\

\noindent\textbf{Data Availability} The datasets generated and analysed during the current study are available as a part of our replication package at the following link: \url{https://doi.org/10.5281/zenodo.21838823} \\

\noindent\textbf{Code Availability} All data and materials that support the findings of this study are available in our replication package at the following link: \url{https://doi.org/10.5281/zenodo.21838823} \\

%% file: 10_appendix.tex
\section*{Appendix}

\begin{table}
\centering
\vspace*{\fill}
\caption{Definition of each stage of the \textit{DS-Pipelines} taxonomy \taxodspipelines.}
\label{tab:dspipelines_taxonomy}
\begin{tabular}{lp{8.7cm}}
\toprule
stage & definition \\
\midrule
Data Acquisition & In the beginning of DS pipeline, data are collected from appropriate sources. Data can be acquired manually or automatically. Data acquisition also involves understanding the nature of the data, collecting relevant data, and integrating available datasets. \\
Data Preparation & Data are generally acquired in a raw format that needs certain preprocessing steps. This involves exploration and filtering, which helps identify the correct data for further processing. Well prepared data reduces the time required for data analysis and contributes to the success of the DS pipeline. \\
Storage & It is important to find an appropriate hardware-software combination to preserve data so that it can be processed efficiently. For example, Miao et al. used graph database system Neo4j [52] to build a collaborative analysis pipeline [49], since Neo4J supports querying graph data properties. \\
Feature Engineering & The entire dataset might not contribute equally to decision making. In this stage, appropriate features that are useful to build the model are identified or constructed. Features that are not readily available in the dataset, require engineering to create them from raw data. \\
Modeling & When data are preprocessed and features are extracted, a model is built to analyze the data. Model building includes model planning, model selection, mining and deriving important properties of data. Appropriate data processing strategies and algorithms are selected to create a good model. \\
Training & For a specific model, we need to train the model with available labeled data. By each training iteration, we optimize the model and try to make it better. The quality of the training dataset contributes to the training accuracy of the model. \\
Evaluation & After training the model, it is tested with a new dataset which has not been used as training data. Also, the model can be evaluated in real-life scenarios and compared with other competing models. Existing metrics are used or new metrics are created to evaluate the model. \\
Prediction & The success of the model depends on how good a model can predict in an unknown setup. After a satisfactory evaluation, we employ the model to solve the problem and see how it works. There are many prediction metrics such as classification accuracy, log-loss, F1-score, to measure the success of the model. \\
Interpretation & The prediction result might not be enough to make a decision. We often need a transformation of the prediction result and post-processing to translate predictions into knowledge. For example, only numerical results do not help much but a good visualization can help to make a decision. \\
Communication & Different components of the DS system might reside in a distributed environment. So, we might need to communicate with the involved parties (e.g., devices, persons, systems) to share and accumulate information. Communication might take place in different geographical locations or the same. \\
Deployment & The built DS solution is installed in its problem domain to serve the application. Over time, the performance of the model is monitored so that the model can be improved to handle new situations. Deployment also includes model maintenance and sending feedback to the model building layer. \\
\bottomrule
\end{tabular}
\vspace*{\fill}
\end{table}

\begin{table}
\centering
\vspace*{\fill}
\caption{Definition of each stage of the \textit{DASWOW} taxonomy \taxodaswow.}
\label{tab:daswow_taxonomy}
\begin{tabular}{lp{8.8cm}}
\toprule
stage & definition \\
\midrule
helper\_functions & Code that is not directly related to the data science activity at hand,  but provides useful scripting functions (e. g. importing or configuring libraries). \\
load\_data & The process of loading a dataset of any type (e.g., .csv, .pkl) into a  Jupyter notebook environment. \\
data\_preprocessing & The process of preparing the dataset(s) for the subsequent analysis. It  includes tasks such as cleaning, instance selection, normalisation, data transformation, and feature selection. \\
data\_exploration & The process of inspecting the content and shape of a dataset to understand the nature and characteristics of the data. Note that it may involve the usage of visualisation techniques but differs in its purpose. \\
modelling & The process of applying statistical models and learning-based algorithms to learn from sample data. \\
evaluation & The process of assessing a model using one/various evaluation metric(s)  such as goodness of fit and accuracy. \\
prediction & The process of applying a model trained on a set of data to other or  newly arriving pieces of data to forecast new values. \\
result\_visualization & The process of obtaining a graphical representation (e.g., tables, plots,  graphs) of a/several measurement(s). \\
save\_results & The process of serialising and storing the data. \\
comment\_only & Lines of comment including commented code. \\
\bottomrule
\end{tabular}
\vspace*{\fill}
\end{table}

\begin{table}
\centering
\vspace*{\fill}
\caption{Definition of each stage of the best performing taxonomy mutation. Highlighted stage corresponds to the mutated headword.}
\label{tab:best_mutated_unified_taxonomy}
\begin{tabular}{lp{8.8cm}}
\toprule
stage & definition \\
\midrule
Data Collection & The process of gathering and importing data from various sources into a data analysis environment for further processing and analysis. \\
Data Preparation & The initial phase of data science where raw data is cleaned, filtered, transformed, and normalized to ensure quality and relevance for analysis. \\
Data Modeling & The process of creating mathematical or computational representations of real-world phenomena using statistical and machine learning techniques to extract insights and make predictions. \\
\textbf{Model Prediction} & The application of a previously trained machine learning model to new, unseen data for the purpose of making predictions or decisions. \\
Model Evaluation & The process of assessing a trained model's performance using specific metrics and comparing it against new datasets or real-world scenarios to determine its effectiveness and reliability. \\
Save Results & The process of serializing and persisting data for future use or analysis. \\
\bottomrule
\end{tabular}
\vspace*{\fill}
\end{table}

\begin{lstlisting}[
    basicstyle=\ttfamily\scriptsize,
    label={lst:system_prompt},
    caption={Template for the system prompt.}
]
```python
{{ all_python_code|join('') }}
```
\end{lstlisting}

\begin{lstlisting}[
    basicstyle=\ttfamily\scriptsize,
    label={lst:user_prompt},
    caption={Template for the user prompt, before applying mutations.}
]
<|task|>

# You are a code classification expert who is an experienced Data Scientist, with years of experience and extensive knowledge about data science and machine learning libraries.
# You have studied the code above, paying attention to its structure and flow.
# Now, you use the following function to classify it.
# Make sure you fully understand each stage definition and consider all of them before making a choice.

# Among the following categories, which one applies to the code snippet
def classify_pipeline_code(code_snippet: str):
    """Classify the code snippet given the following data science lifecycle stages:

    * Data Collection: The process of gathering and importing data from various sources into a data analysis environment for further processing and analysis.
    * Data Preparation: The initial phase of data science where raw data is cleaned, filtered, transformed, and normalized to ensure quality and relevance for analysis.
    * Data Modeling: The process of creating mathematical or computational representations of real-world phenomena using statistical and machine learning techniques to extract insights and make predictions.
    * Model Evaluation: The process of assessing a trained model's performance using specific metrics and comparing it against new datasets or real-world scenarios to determine its effectiveness and reliability.
    * Model Deployment: The application of a previously trained machine learning model to new, unseen data for the purpose of making predictions or decisions.
    * Save Results: The process of serializing and persisting data for future use or analysis.

    If the code snippet does not match any of the stages, returns "Others".

    Args:
        code_snippet (str): the code snippet, extracted from the data science code

    Returns:
        str: the stage corresponding to the code snippet
    """
    if code_snippet.contributes_to("the process of assessing a trained model's performance using specific metrics and comparing it against new datasets or real-world scenarios to determine its effectiveness and reliability"):
        # the code evaluates a model
        return "Model Evaluation"
    if code_snippet.contributes_to("the application of a previously trained machine learning model to new, unseen data for the purpose of making predictions or decisions"):
        # the code makes inference using a model
        # it does not transform data, which could be related to data preparation
        return "Model Deployment"
    if code_snippet.contributes_to("the process of serializing and persisting data for future use or analysis"):
        # the code saves data or prediction results for persistence
        # it does not transform data, which could be related to data preparation
        return "Save Results"
    if code_snippet.contributes_to("gathering and importing data from various sources into a data analysis environment for further processing and analysis"):
        # the code snippet gathers and imports data from various sources into a data analysis environment for further processing and analysis
        return "Data Collection"
    if code_snippet.contributes_to("creating mathematical or computational representations of real-world phenomena using statistical and machine learning techniques to extract insights and make predictions"):
        # the code snippet creates, instantiates, build or train a model
        # it does not transform data, which could be related to Data Preparation
        return "Data Modeling"
    if code_snippet.contributes_to("the initial phase of data science where raw data is cleaned, filtered, transformed, and normalized to ensure quality and relevance for analysis"):
        # the code snippet contributes to the initial phase of data science where raw data is cleaned, filtered, transformed, and normalized to ensure quality and relevance for analysis
        return "Data Preparation"
    else:
        # it is 100% sure the code does not contribute to any other stage
        return "Others"

# For example:

- Example 1: `classify_pipeline_code("df = pd.DataFrame()")` should return `"Data Preparation"`
- Example 2: `classify_pipeline_code("model.fit()")` should return `"Data Modeling"`
- Example 3: `classify_pipeline_code("predictions = model.predict(X_test)")` should return `"Model Deployment"`
- Example 4: `classify_pipeline_code("predictions.reshape(-1)")` should return `"Data Preparation"`
- Example 5: `classify_pipeline_code("Dense(128)")` should return `"Data Modeling"`
- Example 6: `classify_pipeline_code(data.plot(kind='line', legend=None, figsize=(5, 5))")` return `"Model Evaluation"`
- Example 7: `classify_pipeline_code(ax.set_xticks(range(len(farmers)), labels=farmers, rotation=45, ha="right", rotation_mode="anchor")")` return `"Model Evaluation"`
- Example 8: `classify_pipeline_code("im = ax.imshow(harvest)")` return `"Model Evaluation"`

{% if is_multi_class %}
    # You should improve an existing classifier that returned one of the following results:
{% else %}
    # You should improve an existing classifier that returned the following result:
{% endif %}

{{ expected_class }}

# What would be the result for the following code snippet?

{{ python_code_line }}
\end{lstlisting}
\FloatBarrier

\begin{table}
\caption{Statistical significance for all tested SLMs pairwise comparisons with Holm-Bonferroni correction.}
\label{tab:all_slms_comparisons_significance}
\begin{adjustbox}{max width=\textwidth}
\begin{tabular}{lrrrrr}
\toprule
comparison & significant\_difference & odds\_ratio & chi2 & dof & p-corrected \\
\midrule
('Qwen2.5-Coder-7B-Instruct', 'ibm-granite') & True & 6.31E-02 & 7.31E+02 & 1 & 5.42E-192 \\
('Qwen2.5-Coder-7B-Instruct', 'gemma-3-4b-it') & True & 1.88E-01 & 2.84E+02 & 1 & 8.99E-69 \\
('Qwen2.5-Coder-7B-Instruct', 'Phi-3-mini-128k-instruct') & True & 4.91E-02 & 8.24E+02 & 1 & 2.11E-220 \\
('Qwen2.5-Coder-7B-Instruct', 'CodeQwen1.5-7B-Chat') & True & 5.74E-02 & 7.74E+02 & 1 & 1.02E-204 \\
('Qwen2.5-Coder-7B-Instruct', 'Magicoder-S-DS-6.7B') & True & 5.09E-01 & 3.68E+01 & 1 & 3.72E-09 \\
('Qwen2.5-Coder-7B-Instruct', 'Artigenz-Coder-DS-6.7B') & True & 3.31E-01 & 1.04E+02 & 1 & 9.70E-25 \\
('ibm-granite', 'gemma-3-4b-it') & True & 1.11E+01 & 3.43E+02 & 1 & 3.52E-88 \\
('ibm-granite', 'Phi-3-mini-128k-instruct') & True & 3.70E-01 & 3.59E+01 & 1 & 3.72E-09 \\
('ibm-granite', 'CodeQwen1.5-7B-Chat') & True & 2.91E-01 & 2.03E+01 & 1 & 7.51E-06 \\
('ibm-granite', 'Magicoder-S-DS-6.7B') & True & 7.95E+00 & 5.54E+02 & 1 & 5.70E-138 \\
('ibm-granite', 'Artigenz-Coder-DS-6.7B') & True & 7.27E+00 & 4.69E+02 & 1 & 4.99E-116 \\
('gemma-3-4b-it', 'Phi-3-mini-128k-instruct') & True & 8.70E-02 & 4.12E+02 & 1 & 4.04E-106 \\
('gemma-3-4b-it', 'CodeQwen1.5-7B-Chat') & True & 1.05E-01 & 3.65E+02 & 1 & 1.77E-92 \\
('gemma-3-4b-it', 'Magicoder-S-DS-6.7B') & True & 2.36E+00 & 1.22E+02 & 1 & 2.96E-28 \\
('gemma-3-4b-it', 'Artigenz-Coder-DS-6.7B') & True & 1.84E+00 & 6.10E+01 & 1 & 1.77E-14 \\
('Phi-3-mini-128k-instruct', 'CodeQwen1.5-7B-Chat') & True & 1.87E+00 & 1.19E+01 & 1 & 5.27E-04 \\
('Phi-3-mini-128k-instruct', 'Magicoder-S-DS-6.7B') & True & 1.79E+01 & 7.10E+02 & 1 & 2.17E-188 \\
('Phi-3-mini-128k-instruct', 'Artigenz-Coder-DS-6.7B') & True & 1.59E+01 & 6.16E+02 & 1 & 3.29E-162 \\
('CodeQwen1.5-7B-Chat', 'Magicoder-S-DS-6.7B') & True & 1.21E+01 & 6.38E+02 & 1 & 2.56E-164 \\
('CodeQwen1.5-7B-Chat', 'Artigenz-Coder-DS-6.7B') & True & 1.07E+01 & 5.46E+02 & 1 & 3.07E-139 \\
('Magicoder-S-DS-6.7B', 'Artigenz-Coder-DS-6.7B') & True & 1.74E-01 & 6.55E+01 & 1 & 1.26E-16 \\
\bottomrule
\end{tabular}
\end{adjustbox}
\end{table}

\begin{table}
\caption{Statistical values for mutations pairwise comparisons with Holm-Bonferroni correction where no significant difference was observed.}
\label{tab:mutations_comparisons_significance}
\begin{adjustbox}{max width=\textwidth}
\begin{tabular}{lrrrrr}
\toprule
comparison & significant\_difference & odds\_ratio & chi2 & dof & p-corrected \\
\midrule
('mutation\_original', 'mutation\_2') & False & 9.20E-01 & 5.43E-01 & 1 & 1.00E+00 \\
('mutation\_original', 'mutation\_5') & False & 9.50E-01 & 2.61E-01 & 1 & 1.00E+00 \\
('mutation\_1', 'mutation\_3') & False & 7.10E-01 & 3.06E+00 & 1 & 1.00E+00 \\
('mutation\_1', 'mutation\_10') & False & 6.34E-01 & 5.39E+00 & 1 & 4.77E-01 \\
('mutation\_1', 'mutation\_11') & False & 1.31E+00 & 1.77E+00 & 1 & 1.00E+00 \\
('mutation\_1', 'mutation\_12') & False & 1.08E+00 & 8.18E-02 & 1 & 1.00E+00 \\
('mutation\_1', 'mutation\_14') & False & 1.27E+00 & 8.53E-01 & 1 & 1.00E+00 \\
('mutation\_1', 'mutation\_17') & False & 9.79E-01 & 0.00E+00 & 1 & 1.00E+00 \\
('mutation\_1', 'mutation\_18') & False & 5.41E-01 & 7.76E+00 & 1 & 1.52E-01 \\
('mutation\_2', 'mutation\_5') & False & 1.01E+00 & 8.60E-03 & 1 & 1.00E+00 \\
('mutation\_3', 'mutation\_10') & False & 9.25E-01 & 1.62E-01 & 1 & 1.00E+00 \\
('mutation\_3', 'mutation\_11') & False & 1.81E+00 & 9.55E+00 & 1 & 5.81E-02 \\
('mutation\_3', 'mutation\_12') & False & 1.47E+00 & 4.20E+00 & 1 & 8.63E-01 \\
('mutation\_3', 'mutation\_13') & False & 6.86E-01 & 5.59E+00 & 1 & 4.46E-01 \\
('mutation\_3', 'mutation\_14') & False & 1.67E+00 & 6.82E+00 & 1 & 2.53E-01 \\
('mutation\_3', 'mutation\_17') & False & 1.43E+00 & 3.03E+00 & 1 & 1.00E+00 \\
('mutation\_3', 'mutation\_18') & False & 8.86E-01 & 3.71E-01 & 1 & 1.00E+00 \\
('mutation\_4', 'mutation\_6') & False & 8.10E-01 & 2.52E+00 & 1 & 1.00E+00 \\
('mutation\_10', 'mutation\_13') & False & 5.59E-01 & 6.79E+00 & 1 & 2.53E-01 \\
('mutation\_10', 'mutation\_17') & False & 1.74E+00 & 6.19E+00 & 1 & 3.22E-01 \\
('mutation\_10', 'mutation\_18') & False & 9.56E-01 & 1.14E-02 & 1 & 1.00E+00 \\
('mutation\_11', 'mutation\_12') & False & 6.94E-01 & 1.64E+00 & 1 & 1.00E+00 \\
('mutation\_11', 'mutation\_14') & False & 8.70E-01 & 2.91E-01 & 1 & 1.00E+00 \\
('mutation\_11', 'mutation\_17') & False & 5.43E-01 & 4.17E+00 & 1 & 8.63E-01 \\
('mutation\_12', 'mutation\_14') & False & 1.14E+00 & 2.08E-01 & 1 & 1.00E+00 \\
('mutation\_12', 'mutation\_17') & False & 8.53E-01 & 2.54E-01 & 1 & 1.00E+00 \\
('mutation\_13', 'mutation\_15') & False & 5.38E-01 & 6.61E+00 & 1 & 2.61E-01 \\
('mutation\_13', 'mutation\_16') & False & 7.29E-01 & 4.24E+00 & 1 & 8.63E-01 \\
('mutation\_13', 'mutation\_18') & False & 1.63E+00 & 5.29E+00 & 1 & 4.83E-01 \\
('mutation\_14', 'mutation\_17') & False & 7.44E-01 & 1.19E+00 & 1 & 1.00E+00 \\
('mutation\_15', 'mutation\_16') & False & 9.51E-01 & 7.96E-02 & 1 & 1.00E+00 \\
\bottomrule
\end{tabular}
\end{adjustbox}
\end{table}

\begin{figure}[t]
\label{fig:daswow_daswow_confmatrix}
\caption{Normalized multi-label classification matrix of \textit{DASWOW} classification}
\includegraphics[width=11cm]{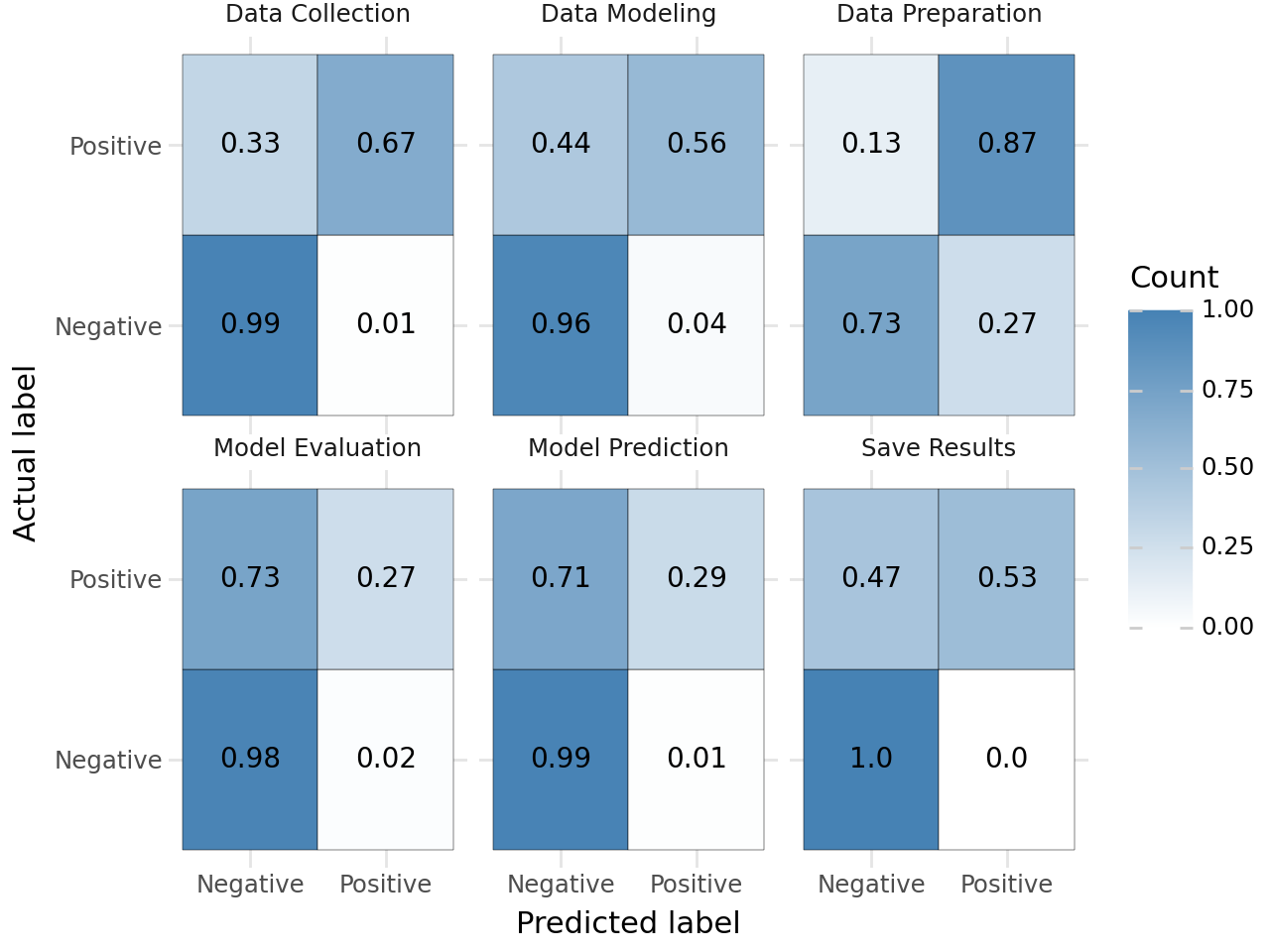}
\centering
\end{figure}

\begin{table}
\caption{Adjusted Pearson residuals for the distribution of pipeline stages. Statistically significant residuals appear in grey.}
\label{tab:adj_pearson_residuals_distrib_stages}
\begin{tabular}{lrrrr}
\toprule
stage & adjusted residual & p-value raw & p-value corrected & significant \\
\midrule
\multicolumn{5}{c}{DASWOW} \\
\midrule
Data Collection & -1.26E+00 & 2.06E-01 & 4.13E-01 & False \\
Data Modeling & -2.16E+00 & 3.08E-02 & 9.23E-02 & False \\
\rowcolor{lightgray} Data Preparation & 9.97E+00 & 2.16E-23 & 1.30E-22 & True \\
\rowcolor{lightgray} Model Evaluation & -7.91E+00 & 2.60E-15 & 1.30E-14 & True \\
\rowcolor{lightgray} Model Prediction & -4.76E+00 & 1.98E-06 & 7.90E-06 & True \\
Save Results & -1.21E+00 & 2.25E-01 & 4.13E-01 & False \\
\midrule
\multicolumn{5}{c}{DSPipelines} \\
\midrule
\rowcolor{lightgray} Data Collection & 7.16E+00 & 7.83E-13 & 4.70E-12 & True \\
\rowcolor{lightgray} Data Modeling & -2.57E+00 & 1.01E-02 & 4.03E-02 & True \\
Data Preparation & -1.06E+00 & 2.91E-01 & 2.91E-01 & False \\
Model Evaluation & -1.74E+00 & 8.27E-02 & 1.95E-01 & False \\
Model Prediction & 1.85E+00 & 6.50E-02 & 1.95E-01 & False \\
\rowcolor{lightgray} Save Results & -3.73E+00 & 1.93E-04 & 9.67E-04 & True \\
\midrule
\multicolumn{5}{c}{\slmbest} \\
\midrule
Data Collection & -5.02E-01 & 6.15E-01 & 9.71E-01 & False \\
\rowcolor{lightgray} Data Modeling & -9.90E+00 & 4.21E-23 & 2.53E-22 & True \\
Data Preparation & -6.97E-01 & 4.86E-01 & 9.71E-01 & False \\
\rowcolor{lightgray} Model Evaluation & 8.93E+00 & 4.15E-19 & 2.08E-18 & True \\
\rowcolor{lightgray} Model Prediction & 4.55E+00 & 5.30E-06 & 2.12E-05 & True \\
Save Results & 2.07E+00 & 3.88E-02 & 1.16E-01 & False \\
\bottomrule
\end{tabular}
\end{table}

%% file: references.bib
@article{hou2024large,
  title={Large language models for software engineering: A systematic literature review},
  author={Hou, Xinyi and Zhao, Yanjie and Liu, Yue and Yang, Zhou and Wang, Kailong and Li, Li and Luo, Xiapu and Lo, David and Grundy, John and Wang, Haoyu},
  journal={ACM Transactions on Software Engineering and Methodology},
  volume={33},
  number={8},
  pages={1--79},
  year={2024},
  publisher={ACM New York, NY}
}

@article{wang2022documentation,
  title={Documentation matters: Human-centered ai system to assist data science code documentation in computational notebooks},
  author={Wang, April Yi and Wang, Dakuo and Drozdal, Jaimie and Muller, Michael and Park, Soya and Weisz, Justin D and Liu, Xuye and Wu, Lingfei and Dugan, Casey},
  journal={ACM Transactions on Computer-Human Interaction},
  volume={29},
  number={2},
  pages={1--33},
  year={2022},
  publisher={ACM New York, NY}
}

@inproceedings{wang2021graph,
  title={Graph-Augmented Code Summarization in Computational Notebooks.},
  author={Wang, April Yi and Wang, Dakuo and Liu, Xuye and Wu, Lingfei and others},
  booktitle={IJCAI},
  pages={5020--5023},
  year={2021}
}

@article{venkatesh2024static,
  title={Static analysis driven enhancements for comprehension in machine learning notebooks},
  author={Venkatesh, Ashwin Prasad Shivarpatna and Sabu, Samkutty and Chekkapalli, Mouli and Wang, Jiawei and Li, Li and Bodden, Eric},
  journal={Empirical Software Engineering},
  volume={29},
  number={5},
  pages={136},
  year={2024},
  publisher={Springer}
}

@inproceedings{venkatesh2023enhancing,
  title={Enhancing comprehension and navigation in jupyter notebooks with static analysis},
  author={Venkatesh, Ashwin Prasad Shivarpatna and Wang, Jiawei and Li, Li and Bodden, Eric},
  booktitle={2023 IEEE international Conference on software analysis, evolution and reengineering (SANER)},
  pages={391--401},
  year={2023},
  organization={IEEE}
}

@article{ramasamy2023workflow,
  title={Workflow analysis of data science code in public GitHub repositories},
  author={Ramasamy, Dhivyabharathi and Sarasua, Cristina and Bacchelli, Alberto and Bernstein, Abraham},
  journal={Empirical Software Engineering},
  volume={28},
  number={1},
  pages={7},
  year={2023},
  publisher={Springer}
}

@article{perez2024flexible,
  title={A Flexible Cell Classification for ML Projects in Jupyter Notebooks},
  author={Perez, Miguel and Aydin, Selin and Lichter, Horst},
  journal={arXiv preprint arXiv:2403.07562},
  year={2024}
}

@inproceedings{biswas2022art,
  title={The art and practice of data science pipelines: A comprehensive study of data science pipelines in theory, in-the-small, and in-the-large},
  author={Biswas, Sumon and Wardat, Mohammad and Rajan, Hridesh},
  booktitle={Proceedings of the 44th International Conference on Software Engineering},
  pages={2091--2103},
  year={2022}
}

@article{patterson2017dataflow,
  title={Dataflow representation of data analyses: Toward a platform for collaborative data science},
  author={Patterson, Evan and McBurney, Robert and Schmidt, Holly and Baldini, Ioana and Mojsilovi{\'c}, Aleksandra and Varshney, Kush R},
  journal={IBM Journal of Research and Development},
  volume={61},
  number={6},
  pages={9--1},
  year={2017},
  publisher={IBM}
}

@book{national2019reproducibility,
  title={Reproducibility and replicability in science},
  author={National Academies of Sciences and Medicine and Policy and Global Affairs and Board on Research Data and Division on Engineering and Physical Sciences and Committee on Applied and Theoretical Statistics and Board on Mathematical Sciences and others},
  year={2019},
  publisher={National Academies Press}
}

@inproceedings{jiang2022elevating,
  title={Elevating jupyter notebook maintenance tooling by identifying and extracting notebook structures},
  author={Jiang, Yuan and K{\"a}stner, Christian and Zhou, Shurui},
  booktitle={2022 IEEE International Conference on Software Maintenance and Evolution (ICSME)},
  pages={399--403},
  year={2022},
  organization={IEEE}
}

@article{zhang2022coral,
  title={Coral: Code representation learning with weakly-supervised transformers for analyzing data analysis},
  author={Zhang, Ge and Merrill, Mike A and Liu, Yang and Heer, Jeffrey and Althoff, Tim},
  journal={EPJ Data Science},
  volume={11},
  number={1},
  pages={14},
  year={2022},
  publisher={Springer Berlin Heidelberg}
}

@inproceedings{garijo2013detecting,
  title={Detecting common scientific workflow fragments using templates and execution provenance},
  author={Garijo, Daniel and Corcho, Oscar and Gil, Yolanda},
  booktitle={Proceedings of the seventh international conference on Knowledge capture},
  pages={33--40},
  year={2013}
}

@inproceedings{rule2018exploration,
  title={Exploration and explanation in computational notebooks},
  author={Rule, Adam and Tabard, Aur{\'e}lien and Hollan, James D},
  booktitle={Proceedings of the 2018 CHI conference on human factors in computing systems},
  pages={1--12},
  year={2018}
}

@article{lozhkov2024starcoder,
  title={Starcoder 2 and the stack v2: The next generation},
  author={Lozhkov, Anton and Li, Raymond and Allal, Loubna Ben and Cassano, Federico and Lamy-Poirier, Joel and Tazi, Nouamane and Tang, Ao and Pykhtar, Dmytro and Liu, Jiawei and Wei, Yuxiang and others},
  journal={arXiv preprint arXiv:2402.19173},
  year={2024}
}

@article{aggarwal2019can,
  title={How can AI automate End-to-End data science?},
  author={Aggarwal, Charu and Bouneffouf, Djallel and Samulowitz, Horst and Buesser, Beat and Hoang, Thanh and Khurana, Udayan and Liu, Sijia and Pedapati, Tejaswini and Ram, Parikshit and Rawat, Ambrish and others},
  journal={arXiv preprint arXiv:1910.14436},
  year={2019}
}

@article{russo2024generative,
  title={Generative ai in software engineering must be human-centered: The copenhagen manifesto},
  author={Russo, Daniel and Baltes, Sebastian and van Berkel, Niels and Avgeriou, Paris and Calefato, Fabio and Cabrero-Daniel, Beatriz and Catolino, Gemma and Cito, J{\"u}rgen and Ernst, Neil and Fritz, Thomas and others},
  journal={Journal of Systems and Software},
  volume={216},
  pages={112115},
  year={2024}
}

@article{schulhoff2024prompt,
  title={The prompt report: a systematic survey of prompt engineering techniques},
  author={Schulhoff, Sander and Ilie, Michael and Balepur, Nishant and Kahadze, Konstantine and Liu, Amanda and Si, Chenglei and Li, Yinheng and Gupta, Aayush and Han, HyoJung and Schulhoff, Sevien and others},
  journal={arXiv preprint arXiv:2406.06608},
  year={2024}
}

@inproceedings{bogner2021characterizing,
  title={Characterizing technical debt and antipatterns in AI-based systems: A systematic mapping study},
  author={Bogner, Justus and Verdecchia, Roberto and Gerostathopoulos, Ilias},
  booktitle={2021 IEEE/ACM International Conference on Technical Debt (TechDebt)},
  pages={64--73},
  year={2021},
  organization={IEEE}
}

@article{redyuk2024assisted,
  title={Assisted design of data science pipelines},
  author={Redyuk, Sergey and Kaoudi, Zoi and Schelter, Sebastian and Markl, Volker},
  journal={The VLDB Journal},
  volume={33},
  number={4},
  pages={1129--1153},
  year={2024},
  publisher={Springer}
}

@inproceedings{brault2023taming,
  title={Taming the Diversity of Computational Notebooks},
  author={Brault, Yann and El Amraoui, Yassine and Blay-Fornarino, Mireille and Collet, Philippe and Jaillet, Florent and Precioso, Fr{\'e}d{\'e}ric},
  booktitle={Proceedings of the 27th ACM International Systems and Software Product Line Conference-Volume A},
  pages={27--33},
  year={2023}
}

@article{martinez2019crisp,
  title={CRISP-DM twenty years later: From data mining processes to data science trajectories},
  author={Mart{\'\i}nez-Plumed, Fernando and Contreras-Ochando, Lidia and Ferri, Cesar and Hern{\'a}ndez-Orallo, Jos{\'e} and Kull, Meelis and Lachiche, Nicolas and Ram{\'\i}rez-Quintana, Mar{\'\i}a Jos{\'e} and Flach, Peter},
  journal={IEEE transactions on knowledge and data engineering},
  volume={33},
  number={8},
  pages={3048--3061},
  year={2019},
  publisher={IEEE}
}

@inproceedings{DMwirth2000crisp,
	title = {{CRISP}-{DM}: {Towards} a standard process model for data mining},
	volume = {1},
	shorttitle = {{CRISP}-{DM}},
	url = {http://www.cs.unibo.it/~danilo.montesi/CBD/Beatriz/10.1.1.198.5133.pdf},
	urldate = {2025-03-12},
	booktitle = {Proceedings of the 4th international conference on the practical applications of knowledge discovery and data mining},
	publisher = {Manchester},
	author = {Wirth, Rüdiger and Hipp, Jochen},
	year = {2000},
	pages = {29--39}
}

@article{stevens2001classification,
  title={A classification of tasks in bioinformatics},
  author={Stevens, Robert and Goble, Carole and Baker, Patricia and Brass, Andy},
  journal={Bioinformatics},
  volume={17},
  number={2},
  pages={180--188},
  year={2001},
  publisher={Oxford University Press}
}

@inproceedings{wassink2009analysing,
  title={Analysing scientific workflows: why workflows not only connect web services},
  author={Wassink, Ingo and Van Der Vet, Paul E and Wolstencroft, Katy and Neerincx, Pieter BT and Roos, Marco and Rauwerda, Han and Breit, Timo M},
  booktitle={2009 Congress on Services-I},
  pages={314--321},
  year={2009},
  organization={IEEE}
}

@article{liu2015survey,
  title={A survey of data-intensive scientific workflow management},
  author={Liu, Ji and Pacitti, Esther and Valduriez, Patrick and Mattoso, Marta},
  journal={Journal of Grid Computing},
  volume={13},
  number={4},
  pages={457--493},
  year={2015},
  publisher={Springer}
}

@article{chen2021evaluating,
  title={Evaluating large language models trained on code},
  author={Chen, Mark},
  journal={arXiv preprint arXiv:2107.03374},
  year={2021}
}

@article{wang2025machine,
  title={Why do Machine Learning Notebooks Crash? An Empirical Study on Public Python Jupyter Notebooks},
  author={Wang, Yiran and Meijer, Willem and Lopez, Jose Antonio Hernandez and Nilsson, Ulf and Varro, Daniel},
  journal={IEEE Transactions on Software Engineering},
  year={2025},
  publisher={IEEE}
}

@inproceedings{islam2019comprehensive,
  title={A comprehensive study on deep learning bug characteristics},
  author={Islam, Md Johirul and Nguyen, Giang and Pan, Rangeet and Rajan, Hridesh},
  booktitle={Proceedings of the 2019 27th ACM joint meeting on european software engineering conference and symposium on the foundations of software engineering},
  pages={510--520},
  year={2019}
}

@article{card2020little,
  title={With little power comes great responsibility},
  author={Card, Dallas and Henderson, Peter and Khandelwal, Urvashi and Jia, Robin and Mahowald, Kyle and Jurafsky, Dan},
  journal={arXiv preprint arXiv:2010.06595},
  year={2020}
}

@book{berman2016essential,
  title={Essential statistics for public managers and policy analysts},
  author={Berman, Evan and Wang, Xiaohu},
  year={2016},
  publisher={Cq Press}
}

@inproceedings{wagner2025towards,
  title={Towards evaluation guidelines for empirical studies involving llms},
  author={Wagner, Stefan and Bar{\'o}n, Marvin Mu{\~n}oz and Falessi, Davide and Baltes, Sebastian},
  booktitle={2025 IEEE/ACM International Workshop on Methodological Issues with Empirical Studies in Software Engineering (WSESE)},
  pages={24--27},
  year={2025},
  organization={IEEE}
}

@inproceedings{boronat2025mdre,
  title={MDRE-LLM: A Tool for Analyzing and Applying LLMs in Software Reverse Engineering},
  author={Boronat, Artur and Mustafa, Jawad},
  booktitle={2025 IEEE International Conference on Software Analysis, Evolution and Reengineering (SANER)},
  pages={850--854},
  year={2025},
  organization={IEEE}
}

@article{siala2025comparison,
  title={A comparison of large language models and model-driven reverse engineering for reverse engineering},
  author={Siala, Hanan and Lano, Kevin},
  journal={Frontiers in Computer Science},
  volume={7},
  pages={1516410},
  year={2025},
  publisher={Frontiers}
}

@article{campanello2025use,
  author = {Victor Campanello and Shariq Shahbaz and Vladislav Indykov and Daniel Strüber},
  title = {On the Use of GPT-4 in the Reverse Engineering of Class Diagrams},
  journal = {Journal of Object Technology},
  volume = {24},
  number = {2},
  issn = {1660-1769},
  year = {2025},
  month = may,
  editor = {},
  note = {The 21st European Conference on Modelling Foundations and Applications (ECMFA 2025)},
  pages = {2:1-14},
  doi = {10.5381/jot.2025.24.2.a14},
  url = {http://www.jot.fm/contents/issue_2025_02/a14.html}
}

@article{landis1977measurement,
  title={The measurement of observer agreement for categorical data},
  author={Landis, J Richard and Koch, Gary G},
  journal={biometrics},
  pages={159--174},
  year={1977},
  publisher={JSTOR}
}

@article{chen2026nlperturbator,
  title={Nlperturbator: Studying the robustness of code llms to natural language variations},
  author={Chen, Junkai and Li, Zhenhao and Hu, Xing and Xia, Xin},
  journal={ACM Transactions on Software Engineering and Methodology},
  volume={35},
  number={4},
  pages={1--20},
  year={2026},
  publisher={ACM New York, NY}
}

@article{bahl1983maximum,
  title={A maximum likelihood approach to continuous speech recognition},
  author={Bahl, Lalit R and Jelinek, Frederick and Mercer, Robert L},
  journal={IEEE transactions on pattern analysis and machine intelligence},
  number={2},
  pages={179--190},
  year={1983},
  publisher={IEEE}
}

@article{bolze2006grid,
  title={Grid'5000: a large scale and highly reconfigurable experimental grid testbed},
  author={Bolze, Rapha{\"e}l and Cappello, Franck and Caron, Eddy and Dayd{\'e}, Michel and Desprez, Fr{\'e}d{\'e}ric and Jeannot, Emmanuel and J{\'e}gou, Yvon and Lanteri, Stephane and Leduc, Julien and Melab, Noredine and others},
  journal={The International Journal of High Performance Computing Applications},
  volume={20},
  number={4},
  pages={481--494},
  year={2006},
  publisher={Sage Publications Sage CA: Thousand Oaks, CA}
}

@inproceedings{brown2020icl,
    author = {Brown, Tom B. and Mann, Benjamin and Ryder, Nick and Subbiah, Melanie and Kaplan, Jared and Dhariwal, Prafulla and Neelakantan, Arvind and Shyam, Pranav and Sastry, Girish and Askell, Amanda and Agarwal, Sandhini and Herbert-Voss, Ariel and Krueger, Gretchen and Henighan, Tom and Child, Rewon and Ramesh, Aditya and Ziegler, Daniel M. and Wu, Jeffrey and Winter, Clemens and Hesse, Christopher and Chen, Mark and Sigler, Eric and Litwin, Mateusz and Gray, Scott and Chess, Benjamin and Clark, Jack and Berner, Christopher and McCandlish, Sam and Radford, Alec and Sutskever, Ilya and Amodei, Dario},
    title = {Language models are few-shot learners},
    year = {2020},
    isbn = {9781713829546},
    publisher = {Curran Associates Inc.},
    address = {Red Hook, NY, USA},
    booktitle = {Proceedings of the 34th International Conference on Neural Information Processing Systems},
    articleno = {159},
    numpages = {25},
    location = {Vancouver, BC, Canada},
    series = {NIPS '20}
}

@inproceedings{puerto2024code,
  title={Code prompting elicits conditional reasoning abilities in text+ code llms},
  author={Puerto, Haritz and Tutek, Martin and Aditya, Somak and Zhu, Xiaodan and Gurevych, Iryna},
  booktitle={Proceedings of the 2024 Conference on Empirical Methods in Natural Language Processing},
  pages={11234--11258},
  year={2024}
}

@article{henderson2020towards,
  title={Towards the systematic reporting of the energy and carbon footprints of machine learning},
  author={Henderson, Peter and Hu, Jieru and Romoff, Joshua and Brunskill, Emma and Jurafsky, Dan and Pineau, Joelle},
  journal={Journal of machine learning research},
  volume={21},
  number={248},
  pages={1--43},
  year={2020}
}

@misc{lacroix2026using,
      title={Using Small Language Models to Reverse-Engineer Machine Learning Pipelines Structures}, 
      author={Nicolas Lacroix and Mireille Blay-Fornarino and Sébastien Mosser and Frederic Precioso},
      year={2026},
      eprint={2601.03988},
      archivePrefix={arXiv},
      primaryClass={cs.SE},
      url={https://arxiv.org/abs/2601.03988}, 
}

@article{zhuo2024bigcodebench,
  title={BigCodeBench: Benchmarking Code Generation with Diverse Function Calls and Complex Instructions},
  author={Zhuo, Terry Yue and Vu, Minh Chien and Chim, Jenny and Hu, Han and Yu, Wenhao and Widyasari, Ratnadira and Yusuf, Imam Nur Bani and Zhan, Haolan and He, Junda and Paul, Indraneil and others},
  journal={arXiv preprint arXiv:2406.15877},
  year={2024}
}

@inproceedings{evalplus,
  title = {Is Your Code Generated by Chat{GPT} Really Correct? Rigorous Evaluation of Large Language Models for Code Generation},
  author = {Liu, Jiawei and Xia, Chunqiu Steven and Wang, Yuyao and Zhang, Lingming},
  booktitle = {Thirty-seventh Conference on Neural Information Processing Systems},
  year = {2023},
  url = {https://openreview.net/forum?id=1qvx610Cu7},
}

@misc{ramasamy_2025_17638925,
  author       = {Ramasamy, Dhivyabharathi and
                  Sarasua, Cristina and
                  Bacchelli, Alberto and
                  Bernstein, Abraham},
  title        = {DASWOW Jupyter Notebooks Subset},
  month        = nov,
  year         = 2025,
  publisher    = {Zenodo},
  doi          = {10.5281/zenodo.17638925},
  url          = {https://doi.org/10.5281/zenodo.17638925},
}

@article{holm1979simple,
  title={A simple sequentially rejective multiple test procedure},
  author={Holm, Sture},
  journal={Scandinavian journal of statistics},
  pages={65--70},
  year={1979},
  publisher={JSTOR}
}

@book{agresti2013categorical,
  title={Categorical data analysis},
  author={Agresti, Alan},
  year={2013},
  publisher={John Wiley \& Sons}
}

@article{lacroixmodel,
  author = {Nicolas Lacroix and Mireille Blay-Fornarino and Philippe Collet and Frédéric Precioso and Sébastien Mosser},
  title = {A Model-Driven Approach To Support The Understanding Of Machine Learning Pipelines},
  journal = {Journal of Object Technology},
  volume = {25},
  number = {3},
  issn = {1660-1769},
  year = {2026},
  editor = {},
  pages = {3:127-140},
  doi = {10.5381/jot.2026.25.3.a10},
  url = {http://www.jot.fm/contents/issue_2026_03/a10.html}
}

@inproceedings{2969442.2969519,
author = {Sculley, D. and Holt, Gary and Golovin, Daniel and Davydov, Eugene and Phillips, Todd and Ebner, Dietmar and Chaudhary, Vinay and Young, Michael and Crespo, Jean-Francois and Dennison, Dan},
title = {Hidden technical debt in Machine learning systems},
year = {2015},
publisher = {MIT Press},
address = {Cambridge, MA, USA},
booktitle = {Proceedings of the 29th International Conference on Neural Information Processing Systems - Volume 2},
pages = {2503–2511},
numpages = {9},
location = {Montreal, Canada},
series = {NIPS'15}
}

@article{granger2021jupyter,
  title={Jupyter: Thinking and storytelling with code and data},
  author={Granger, Brian E and P{\'e}rez, Fernando},
  journal={Computing in Science \& Engineering},
  volume={23},
  number={2},
  pages={7--14},
  year={2021},
  publisher={IEEE}
}

@inproceedings{lau2020design,
  title={The design space of computational notebooks: An analysis of 60 systems in academia and industry},
  author={Lau, Sam and Drosos, Ian and Markel, Julia M and Guo, Philip J},
  booktitle={2020 IEEE Symposium on Visual Languages and Human-Centric Computing (VL/HCC)},
  pages={1--11},
  year={2020},
  organization={IEEE}
}

@article{silla2011survey,
  title={A survey of hierarchical classification across different application domains},
  author={Silla Jr, Carlos N and Freitas, Alex A},
  journal={Data mining and knowledge discovery},
  volume={22},
  number={1},
  pages={31--72},
  year={2011},
  publisher={Springer}
}

@article{hindle2016naturalness,
  title={On the naturalness of software},
  author={Hindle, Abram and Barr, Earl T and Gabel, Mark and Su, Zhendong and Devanbu, Premkumar},
  journal={Communications of the ACM},
  volume={59},
  number={5},
  pages={122--131},
  year={2016},
  publisher={ACM New York, NY, USA}
}

@inproceedings{lea2025rex86,
  title={Rex86: A local large language model for assisting in x86 assembly reverse engineering},
  author={Lea, Darrin and Ghawaly, James and Richard, Golden and Ali-Gombe, Aisha and Case, Andrew},
  booktitle={2025 IEEE Annual Computer Security Applications Conference (ACSAC)},
  pages={108--122},
  year={2025},
  organization={IEEE}
}

@inproceedings{ouf2025reverse,
  title={Reverse Engineering User Stories from Code using Large Language Models},
  author={Ouf, Mohamed and Li, Haoyu and Zhang, Michael and Guizani, Mariam},
  booktitle={2025 IEEE International Conference on Collaborative Advances in Software and COmputiNg (CASCON)},
  pages={504--509},
  year={2025},
  organization={IEEE}
}

@inproceedings{liu2024we,
  title={" we need structured output": Towards user-centered constraints on large language model output},
  author={Liu, Michael Xieyang and Liu, Frederick and Fiannaca, Alexander J and Koo, Terry and Dixon, Lucas and Terry, Michael and Cai, Carrie J},
  booktitle={Extended Abstracts of the CHI Conference on Human Factors in Computing Systems},
  pages={1--9},
  year={2024}
}

@article{wei2022chain,
  title={Chain-of-thought prompting elicits reasoning in large language models},
  author={Wei, Jason and Wang, Xuezhi and Schuurmans, Dale and Bosma, Maarten and Xia, Fei and Chi, Ed and Le, Quoc V and Zhou, Denny and others},
  journal={Advances in neural information processing systems},
  volume={35},
  pages={24824--24837},
  year={2022}
}

@book{cohen2013statistical,
  title={Statistical power analysis for the behavioral sciences},
  author={Cohen, Jacob},
  year={2013},
  publisher={routledge}
}

@article{lakens2022sample,
  title={Sample size justification},
  author={Lakens, Dani{\"e}l},
  journal={Collabra: psychology},
  volume={8},
  number={1},
  pages={33267},
  year={2022},
  publisher={University of California Press}
}

@article{liu2024lost,
  title={Lost in the middle: How language models use long contexts},
  author={Liu, Nelson F and Lin, Kevin and Hewitt, John and Paranjape, Ashwin and Bevilacqua, Michele and Petroni, Fabio and Liang, Percy},
  journal={Transactions of the association for computational linguistics},
  volume={12},
  pages={157--173},
  year={2024}
}
